\documentclass[12pt,a4paper]{article}
\usepackage{bm}
\usepackage{amsmath}
\usepackage{amssymb}

\newcommand{\citen}{\cite}

\makeatletter
\@addtoreset{equation}{section}

\makeatother

\begin{document}

\begin{flushright}
  OCU-PHYS-191 \\
  AP-GR-7
\end{flushright}

\begin{center}\LARGE
  Energy-Momentum and Angular Momentum Carried by Gravitational
  Waves in Extended New General Relativity
\end{center}
\begin{center}\large
  Eisaku \textsc{Sakane}\footnote{%
    E-mail: \texttt{sakane@sci.osaka-cu.ac.jp}}
  and
  Toshiharu \textsc{Kawai}\footnote{E-mail: 
    \texttt{kawai@sci.osaka-cu.ac.jp}}${}^{,}${}\footnote{%
    Present address: 6-5-5, Onodai,
    Osakasayama, Osaka 589-0023, Japan.} \\
  \vspace{1ex}
  \textit{Department of Physics, Graduate School of Science,} \\
  \textit{Osaka City University, Osaka 558-8585, Japan}
\end{center}

\date{\today}

\begin{abstract}
  In an extended, new form of general relativity, which is a
  teleparallel theory of gravity, we examine the energy-momentum and
  angular momentum carried by gravitational wave radiated from
  Newtonian point masses in a weak-field approximation.
  The resulting wave form is identical to the corresponding wave form
  in general relativity, which is consistent with previous results in
  teleparallel theory.
  The expression for the dynamical energy-momentum density is
  identical to that for the canonical energy-momentum density in
  general relativity up to leading order terms on the boundary of a
  large sphere including the gravitational source, and the loss of
  dynamical energy-momentum, which is the generator of \emph{internal}
  translations, is the same as that of the canonical energy-momentum
  in general relativity.
  Under certain asymptotic conditions for a non-dynamical Higgs-type
  field $\psi^{k}$, the loss of \lq\lq{}spin'' angular
  momentum, which is the generator of \emph{internal} $SL(2,C)$
  transformations, is the same as that of angular momentum in general
  relativity, and the losses of canonical energy-momentum and 
  orbital angular momentum, which constitute the generator of
  Poincar\'{e} \emph{coordinate} transformations, are vanishing.
  The results indicate that our definitions of the dynamical
  energy-momentum and angular momentum densities in this extended new
  general relativity work well for gravitational wave radiations, and
  the extended new general relativity accounts for the
  Hulse-Taylor measurement of the pulsar PSR1913+16.
\end{abstract}

\section{Introduction}
\label{sec:introduction}

General relativity (GR) is a standard theory of gravity which has
passed all the observational tests so far carried out, and it
constitutes, together with quantum field theory, a basic framework of
modern theoretical physics.  In GR, however, it is usually
asserted~\cite{misner.thorne.wheeler} that well-behaved
energy-momentum and angular momentum densities cannot be defined in
general for a gravitational field.  For a restricted class of systems
including asymptotically flat space-time, there exist tensor densities
whose integrals over the cross section of the null infinity give the
energy-momentum and angular momentum of the system in
question~\cite{geroch.winicour:1981}.

There are many theories~\cite{hehl.mccrea.mielke.neeman:1995} that are
potential alternatives to GR, including the
$\overline{\mbox{Poincar\'{e}}}$ gauge theory
(\={P}GT)~\cite{kawai:1986} and extended new general relativity
(ENGR)~\cite{kawai.toma:1991}. \={P}GT is formulated on the basis of
the principal fiber bundle over the space-time possessing the covering
group $\bar{P}_{0}$ of the proper orthochronous Poincar\'{e} group as
the structure group, following the standard geometric formulation of
Yang-Mills theories as closely as possible.  ENGR is formulated as the
teleparallel limit of \={P}GT\@.  The dynamical energy-momentum and
\lq\lq{}spin'' angular momentum densities of gravitational and matter
fields are all space-time vector densities, and their integrations
over an arbitrary space-like surface $\sigma$ are well-defined for any
coordinate system employed~\cite{kawai:2000}.

For asymptotically flat space-time whose vierbeins satisfy certain
asymptotic conditions,\footnote{With regard to the situation in ENGR,
  see Eqs.~(\ref{eq:asymptotic_vierbein}) and
  (\ref{eq:asymptotic_vierbein.ii}) in \S\ref{sec:review.ENGR}.}\ 
the integration of the dynamical energy-momentum density over $\sigma$
is the generator of \emph{internal} translations and gives the total
energy-momentum of the system. Also, the integration of the
\lq\lq{}spin'' angular momentum density over $\sigma$ is the generator
of \emph{internal} $SL(2,C)$-transformations and gives the
\emph{total} ($=$\emph{spin}$+$\emph{orbital}) angular momentum in
both theories.  This holds in \={P}GT when the Higgs-type field
$\psi^{k}$ satisfies the asymptotic condition
\begin{math}
  \psi^{k} = 
  e^{(0)k}_{\hspace{2.8ex}\mu}x^{\mu} + \psi^{(0)k} + O(1/r^{\beta})
\end{math}
with constants
\begin{math}
  e^{(0)k}_{\hspace{2.8ex}\mu}, \psi^{(0)k}
\end{math}~\cite{kawai:1988,kawai.saitoh:1989a,kawai.saitoh:1989b}
and in ENGR when this asymptotic condition and certain other
additional conditions are satisfied\cite{kawai.toma:1991}.  These
theories describe within the uncertainties all the observed
gravitational phenomena when the parameters in the gravitational
Lagrangian densities satisfy certain conditions.

Direct observation of gravitational waves is one of the most
challenging problems in present day gravitational physics. Several
projects designed for this purpose are now being carried out, and
gravitational radiations from various possible sources have been
investigated theoretically, mainly on the basis of GR\@.  However,
also in classes of teleparallel theories of gravity, the form of
gravitational waves is known to be identical to that of GR in
post-Newtonian approximations~\cite{schweizer.straumann:1979,%
schweizer.straumann.wipf:1980}.
  
For the case in which a gravitational wave is radiated, however, the
asymptotic behavior of vierbeins is different from that considered in
Ref.~\citen{kawai.toma:1991} in general, and the question of whether
our definitions of the energy-momentum and angular momentum densities
work well in this case should also be answered.

The purpose of the present paper is to examine, in a weak-field
approximation, the energy-momentum and angular momentum carried by
gravitational waves radiated from Newtonian point masses. In
\S\ref{sec:framework}, the basic framework of ENGR is briefly
summarized as preparation for later discussion. In
\S\ref{sec:weak_field_approximation}, the forms of the gravitational
field equations and the dynamical energy-momentum density
\begin{math}
  {}^{G}\bm{T}_{k}^{\hspace{0.7ex}\mu}
\end{math}
of the gravitational field are given in the weak-field approximation.
For plane wave solutions of the linearized homogeneous equations of a
gravitational field, we give the average of
\begin{math}
  {}^{G}\bm{T}_{k}^{\hspace{0.7ex}\mu}
\end{math}
over a space-time region much larger than the inverse of the absolute
value of the three-dimensional wave number vector. In
\S\ref{sec:quadrupole}, the quadrupole radiation formula for a
gravitational wave emitted from a system of Newtonian point masses is
obtained. In \S\ref{sec:emission_rates}, we examine the emission rates of
the dynamical energy-momentum and the angular momentum for two
types of the asymptotic form of a Higgs-type field. Further, the
emission rates of the canonical energy-momentum and the
\lq\lq{}extended orbital angular momentum'' are examined. Finally, in
\S\ref{sec:summary}, we give a summary and discussion.

\section{Basic framework of extended new general relativity}
\label{sec:framework}

\subsection{$\overline{\mbox{Poincar\'{e}}}$ gauge theory}
\label{sec:review.PGT}

We first give the outline of \={P}GT, because ENGR is formulated as a
reduction of this theory.

\={P}GT is formulated on the basis of the principal fiber bundle
$\mathcal{P}$ over the space-time $M$ possessing the covering group
$\bar{P}_{0}$ of the proper orthochronous Poincar\'{e} group as the
structure group. The space-time $M$ is assumed to be a noncompact
four-dimensional differentiable manifold with a countable base. The
bundle $\mathcal{P}$ admits a connection $\Gamma$, the translational
and rotational parts of whose coefficients will be written
\begin{math}
  A^{k}_{\hspace{0.7ex}\mu}
\end{math}
and
\begin{math}
  A^{k}_{\hspace{0.7ex}l\mu}
\end{math},
respectively. The fundamental field variables are
\begin{math}
  A^{k}_{\hspace{0.7ex}\mu}
\end{math}
and
\begin{math}
  A^{k}_{\hspace{0.7ex}l\mu},
\end{math}
the Higgs-type field is
\begin{math}
  \psi = \{\psi^{k}\},
\end{math}
and the matter field is
\begin{math}
  \phi = \{\phi^{A} | A = 1, 2, \ldots, N\}
\end{math}.\footnote{Unless otherwise stated, we use the following
  conventions for indices. Letters from the middle part of the Greek
  alphabet, $\lambda$, $\mu$, $\nu$, $\ldots$, and from the middle
  part of the Latin alphabet, $k$, $l$, $m$, $\ldots$, take the values
  0, 1, 2 and 3.  The capital letters $A$ and $B$ are used as indices
  for components of the field $\phi$, and $N$ denotes the dimension of
  the representation $\rho$.} \ These fields transform according
to\footnote{For the function $f$ on $M$, we define $f_{,\mu}
  \stackrel{\mathrm{def}}{=} \partial f / \partial x^{\mu}$.}
\begin{subequations}
  \begin{align}
    \psi^{\prime k} &=
    (\Lambda(a^{-1}))^{k}_{\hspace{0.7ex}m}(\psi^{m} - t^{m}), \\
    A^{\prime k}_{\hspace{1.0ex}\mu} &=
    (\Lambda(a^{- 1}))^{k}_{\hspace{0.7ex}m}(
      A^{m}_{\hspace{0.7ex}\mu} +
      t^{m}_{\hspace{0.7ex},\mu} + 
      A^{m}_{\hspace{0.7ex}n\mu}t^{n}), \\
    A^{\prime k}_{\hspace{1.0ex}l\mu} &=
    (\Lambda(a^{-1}))^{k}_{\hspace{0.7ex}m}
    A^{m}_{\hspace{0.7ex}n\mu} (\Lambda(a))^{n}_{\hspace{0.7ex}l} +
    (\Lambda(a^{-1}))^{k}_{\hspace{0.7ex}m}
    (\Lambda(a))^{m}_{\hspace{0.7ex}l,\mu} \;, \\
    \phi^{\prime A} &=
    [\rho((t, a)^{-1})]^{A}_{\hspace{0.7ex}B}\phi^{B} \;,
  \end{align}
\end{subequations}
under the $\overline{\mbox{Poincar\'{e}}}$ gauge transformation
\begin{align}
  \label{eq:bar_poincare_gauge_transformation}
  \sigma'(x) &= \sigma(x) \cdot [t(x), a(x)] \;, \nonumber \\
  t(x) &\in T^{4}, \quad a(x) \in SL(2,C).
\end{align}
Here, $\Lambda$ is the covering map from $SL(2,C)$ to the proper
orthochronous Lorentz group, and $\rho$ denotes the representation of
the $\overline{\mbox{Poincar\'{e}}}$ group to which the field
$\phi^{A}$ belongs.  Also, $\sigma$ and $\sigma'$ stand for local
cross sections of $\mathcal{P}$.  The dual components
\begin{math}
  e^{k}_{\hspace{0.7ex}\mu}
\end{math}
of the vierbein fields
\begin{math}
  e^{\mu}_{\hspace{0.7ex}k} \partial / \partial x^{\mu}
\end{math}
are related to the field $\psi^{k}$ and the gauge potentials
\begin{math}
  A^{k}_{\hspace{0.7ex}\mu}
\end{math}
and
\begin{math}
  A^{k}_{\hspace{0.7ex}l\mu}
\end{math}
through the relation
\begin{equation}
  e^{k}_{\hspace{0.7ex}\mu} =
  \psi^{k}_{\hspace{0.7ex},\mu} +
  A^{k}_{\hspace{0.7ex}l\mu} \psi^{l} +
  A^{k}_{\hspace{0.7ex}\mu} \;,
\end{equation}
and these transform according to
\begin{equation}
  e^{\prime k}_{\hspace{1.0ex}\mu} =
  (\Lambda(a^{- 1}))^{k}_{\hspace{0.7ex}l}
  e^{l}_{\hspace{0.7ex}\mu} \;,
\end{equation}
under the transformation (\ref{eq:bar_poincare_gauge_transformation}).
Also, they are related to the metric
\begin{math}
  g_{\mu\nu} dx^{\mu} \otimes dx^{\nu}
\end{math}
of $M$ through the relation
\begin{equation}
  g_{\mu\nu} =
  e^{k}_{\hspace{0.7ex}\mu} \eta_{kl} e^{l}_{\hspace{0.7ex}\nu} \;,
\end{equation}
with 
\begin{math}
  (\eta_{kl}) \stackrel{\mathrm{def}}{=} \textrm{diag}(-1,1,1,1)
\end{math}.

The field strengths
\begin{math}
  R^{k}_{\hspace{0.7ex}l\mu\nu},
  R^{k}_{\hspace{0.7ex}\mu\nu}
\end{math}
and
\begin{math}
  T^{k}_{\hspace{0.7ex}\mu\nu}
\end{math}
of
\begin{math}
  A^{k}_{\hspace{0.7ex}l\mu}, A^{k}_{\hspace{0.7ex}\mu}
\end{math}
and
\begin{math}
  e^{k}_{\hspace{0.7ex}\mu}
\end{math}
are given by\footnote{We define 
\[
  A_{\ldots [ \mu \ldots \nu ] \ldots} \stackrel{\mathrm{def}}{=}
  \frac{1}{2}(A_{\ldots \mu \ldots \nu \ldots} -
    A_{\ldots \nu \ldots \mu \ldots}),
\]
\[
  A_{\ldots ( \mu \ldots \nu ) \ldots} \stackrel{\mathrm{def}}{=}
  \frac{1}{2}(A_{\ldots \mu \ldots \nu \ldots} +
    A_{\ldots \nu \ldots \mu \ldots}).
\]}
\begin{subequations}
  \begin{align}
    R^{k}_{\hspace{0.7ex}l\mu\nu} &\stackrel{\mathrm{def}}{=}
    2(A^{k}_{\hspace{0.7ex}l[\nu,\mu]} +
    A^{k}_{\hspace{0.7ex}m[\mu}A^{m}_{\hspace{0.7ex}l\nu]}), \\
    R^{k}_{\hspace{0.7ex}\mu\nu} &\stackrel{\mathrm{def}}{=}
    2(A^{k}_{\hspace{0.7ex}[\nu,\mu]} +
    A^{k}_{\hspace{0.7ex}l[\mu}A^{l}_{\hspace{0.7ex}\nu]}), \\
    T^{k}_{\hspace{0.7ex}\mu\nu} &\stackrel{\mathrm{def}}{=}
    2(e^{k}_{\hspace{0.7ex}[\nu,\mu]} +
    A^{k}_{\hspace{0.7ex}l[\mu}e^{l}_{\hspace{0.7ex}\nu]}),
  \end{align}
\end{subequations}
respectively, and we have the relation
\begin{equation}
  \label{eq:relations_field_strength}
  T^{k}_{\hspace{0.7ex}\mu\nu} =
  R^{k}_{\hspace{0.7ex}\mu\nu} +
  R^{k}_{\hspace{0.7ex}l\mu\nu} \psi^{l} \;.
\end{equation}
The field strengths
\begin{math}
  T^{k}_{\hspace{0.7ex}\mu\nu}
\end{math}
and
\begin{math}
  R^{k}_{\hspace{0.7ex}l\mu\nu}
\end{math}
are both invariant under \emph{internal} translations.

There is a 2 to 1 bundle homomorphism $F$ from $\mathcal{P}$ to the
affine frame bundle ${\mathcal A}(M)$ over $M$, and there exist an
extended spinor structure and a spinor structure associated with
it~\cite{kawai:1989}. The space-time $M$ is orientable, which follows
from its assumed noncompactness and the fact that $M$ has a spinor
structure.

The affine frame bundle $\mathcal{A}(M)$ admits a connection
$\Gamma_{A}$. The $T^{4}$ part
\begin{math}
  \Gamma^{\mu}_{\hspace{0.7ex}\nu}
\end{math}
and the $GL(4,R)$ part
\begin{math}
  \Gamma^{\lambda}_{\hspace{0.7ex}\mu\nu}
\end{math}
of its connection coefficients are related to
\begin{math}
  A^{k}_{\hspace{0.7ex}l\mu}
\end{math}
and
\begin{math}
  e^{k}_{\hspace{0.7ex}\mu}
\end{math}
through the relations
\begin{subequations}
  \label{eq:relations_connections}
  \begin{align}
    \Gamma^{\mu}_{\hspace{0.7ex}\nu} &=
    \delta^{\mu}_{\hspace{0.7ex}\nu} \;, \\
    A^{k}_{\hspace{0.7ex}l\mu} &=
    e^{k}_{\hspace{0.7ex}\lambda} e^{\nu}_{\hspace{0.7ex}l}
    \Gamma^{\lambda}_{\hspace{0.7ex}\nu\mu} +
    e^{k}_{\hspace{0.7ex}\nu} e^{\nu}_{\hspace{0.7ex}l,\mu} \;,
  \end{align}
\end{subequations}
by the requirement that $F$ maps the connection $\Gamma$ into
$\Gamma_{A}$, and the space-time $M$ is of the Riemann-Cartan type.

The torsion is given by
\begin{equation}
  T^{\lambda}_{\hspace{0.7ex}\mu\nu} \stackrel{\mathrm{def}}{=}
  2\Gamma^{\lambda}_{\hspace{0.7ex}[\nu\mu]} \;,
\end{equation}
and the $T^{4}$ and $GL(4,R)$ parts of the curvature are given by
\begin{align}
  R^{\lambda}_{\hspace{0.7ex}\mu\nu} &=
  2(\Gamma^{\lambda}_{\hspace{0.7ex}[\nu,\mu]} +
  \Gamma^{\lambda}_{\hspace{0.7ex}\rho[\mu} 
    \Gamma^{\rho}_{\hspace{0.7ex}\nu]}), \\
  R^{\lambda}_{\hspace{0.7ex}\rho\mu\nu} &=
  2(\Gamma^{\lambda}_{\hspace{0.7ex}\rho[\nu,\mu]} +
  \Gamma^{\lambda}_{\hspace{0.7ex}\tau[\mu}
    \Gamma^{\tau}_{\hspace{0.7ex}\rho\nu]}),
\end{align}
respectively. Then, we have the relations
\begin{align}
  T^{k}_{\hspace{0.7ex}\mu\nu} &=
  e^{k}_{\hspace{0.7ex}\lambda}T^{\lambda}_{\hspace{0.7ex}\mu\nu} = 
  e^{k}_{\hspace{0.7ex}\lambda}R^{\lambda}_{\hspace{0.7ex}\mu\nu} \;,
  \\
  R^{k}_{\hspace{0.7ex}l\mu\nu} &=
  e^{k}_{\hspace{0.7ex}\lambda}e^{\rho}_{\hspace{0.7ex}l}
    R^{\lambda}_{\hspace{0.7ex}\rho\mu\nu} \;,
\end{align}
which follow from Eq.~(\ref{eq:relations_connections}).

The covariant derivative of the matter field $\phi$ takes the form
\begin{align}
  D_{k}\phi^{A} &=
  e^{\mu}_{\hspace{0.7ex}k} D_{\mu}\phi^{A} \;, \\
  D_{\mu}\phi^{A} &\stackrel{\mathrm{def}}{=}
  \partial_{\mu}\phi^{A} +
  \frac{i}{2}A^{kl}_{\hspace{1.4ex}\mu}(M_{kl}\phi)^{A} +
  iA^{k}_{\hspace{0.7ex}\mu}(P_{k}\phi)^{A} \;,
\end{align}
where $M_{kl}$ and $P_{k}$ are representation matrices of the standard
basis of the Lie algebra of the group $\bar{P}_{0}$:
\begin{math}
  M_{kl} = - i \rho_{*}(\bar{M}_{kl}),
  P_{k} = - i \rho_{*}(\bar{P}_{k}).
\end{math}
The matrix $P_{k}$ represents the \lq\lq{}intrinsic energy-momentum''
of the field $\phi^{A}$~\cite{kawai:1989}, and it is vanishing for
all observed fields.

\subsection{Extended new general relativity}
\label{sec:review.ENGR}

In \={P}GT, we consider the case in which the field strength
\begin{math}
  R^{kl}_{\hspace{1.4ex}\mu\nu}
\end{math}
vanishes identically,
\begin{equation}
  \label{eq:teleparallel_limit}
  R^{kl}_{\hspace{1.4ex}\mu\nu} \equiv 0.
\end{equation}
Thus, the curvature
\begin{math}
  R^{\lambda}_{\hspace{0.7ex}\rho\mu\nu}
\end{math}
vanishes, and we have a teleparallel theory.

By choosing the $SL(2,C)$-gauge such that
\begin{equation}
  \label{eq:lorentz_gauge_fixing}
  A^{kl}_{\hspace{1.4ex}\mu} \equiv 0,
\end{equation}
the following reduced expressions are obtained:
\begin{align}
  \label{eq:engr_relations_vierbein_translation_gauge}
  e^{k}_{\hspace{0.7ex}\mu} &=
  \psi^{k}_{\hspace{0.7ex},\mu} + 
  A^{k}_{\hspace{0.7ex}\mu} \;, \\
  \label{eq:engr_affine_connection_coefficients}
  \Gamma^{\lambda}_{\hspace{0.7ex}\mu\nu} &=
  e^{\lambda}_{\hspace{0.7ex}k}e^{k}_{\hspace{0.7ex}\mu,\nu} \;, \\
  \label{eq:engr_covariant_derivative_latin}
  D_{k}\phi^{A} &=
  e^{\mu}_{\hspace{0.7ex}k}D_{\mu}\phi^{A} \;, \\
  \label{eq:engr_covariant_derivative_greek}
  D_{\mu}\phi^{A} &=
  \partial_{\mu}\phi^{A} +
  i A^{k}_{\hspace{0.7ex}\mu}(P_{k}\phi)^{A} \;.
\end{align}

The Lagrangian takes the form\footnote{The field components
  $e^{k}_{\hspace{0.7ex}\mu}$ and $e^{\mu}_{\hspace{0.7ex}k}$ are used
  to convert Latin and Greek indices.  Also, the raising and lowering
  of the indices $k$, $l$, $m$, $\ldots$ are accomplished through use
  of $(\eta^{kl}) = (\eta_{kl})^{- 1}$ and $(\eta_{kl})$.}
\begin{equation}
  \label{eq:engr_total_lagrangian}
  L = L^{T}(T_{klm}) + 
  L^{M}(e^{k}_{\hspace{0.7ex}\mu},
    \psi^{k}, D_{k}\phi^{A}, \phi^{A}),
\end{equation}
where $L^{M}$ is the Lagrangian of the matter field $\phi^{A}$ and
$L^{T}$ is the gravitational Lagrangian.  We impose the following
requirements: (R.i) $L$ is invariant under the transformation
(\ref{eq:bar_poincare_gauge_transformation}) with arbitrary functions
$t^{k}$ and an arbitrary \emph{constant} element $a$ of $SL(2,C)$;
(R.ii) The functional $L$ is a scalar field on $M$.

The gravitational Lagrangian~\cite{hayashi.shirafuji:1979}
\begin{equation}
  \label{eq:engr_gravitational_lagrangian}
  L^{T} \stackrel{\mathrm{def}}{=}
  c_{1} t^{klm}t_{klm} + c_{2} v^{k}v_{k} + c_{3}a^{k}a_{k}
\end{equation}
satisfies these requirements, where $c_{1}$, $c_{2}$ and $c_{3}$
are real constants. The quantities $t_{klm}$, $v_{k}$ and $a_{k}$ are
the irreducible components of $T_{klm}$ defined by
\begin{align}
  \label{eq:irreducible_component_torsion_tensor_part}
  t_{klm} &\stackrel{\mathrm{def}}{=}
  \frac{1}{2}(T_{klm} + T_{lkm}) +
  \frac{1}{6}(\eta_{mk}v_{l} + \eta_{ml}v_{k}) -
  \frac{1}{3}\eta_{kl}v_{m} \;, \\
  \label{eq:irreducible_component_torsion_vector_part}
  v_{k} &\stackrel{\mathrm{def}}{=}
  T^{l}_{~lk} \;, \\
  \label{eq:irreducible_component_torsion_axial_vector_part}
  a_{k} &\stackrel{\mathrm{def}}{=}
  \frac{1}{6}\epsilon_{klmn}T^{lmn} \;,
\end{align}
where the symbol $\epsilon_{klmn}$ represents the Levi-Civita tensor,
with
\begin{math}
  \epsilon_{(0)(1)(2)(3)} = - 1.
\end{math}\footnote{\label{footnote:distinction_between_indices} Latin
  indices are put in parentheses to distinguish them from Greek
  indices.}

If the parameters $c_{1}$, $c_{2}$ and $c_{3}$ satisfy
\begin{equation}
  \label{eq:teleparallel_equivarence}
  c_{1} = - c_{2} = \frac{4}{9}c_{3} = - \frac{1}{3\kappa} \;,
\end{equation}
where $\kappa$ is the Einstein gravitational constant,
\begin{math}
  \kappa \stackrel{\mathrm{def}}{=} 8\pi G / c^{4},
\end{math}\footnote{$G$ and $c$ stand for the Newtonian gravitational
  constant and the light velocity in vacuum, respectively.}\
the gravitational part of the action integral is equal to the
Einstein-Hilbert action integral, namely~\cite{hayashi.shirafuji:1979}
\begin{equation}
  \int d^{4}x \sqrt{- g} L^{T} =
  \int d^{4}x \frac{1}{2\kappa} \sqrt{- g}R(\{\}),
\end{equation}
where $R(\{\})$ denotes the Riemann-Christoffel scalar curvature.
Here, we should mention that even in the case that the condition
(\ref{eq:teleparallel_equivarence}) is satisfied, our theory does not
reduce to GR, because the couplings of matter fields (the spinor
field, for example) with the gravitational field are different from
those in GR.

In defining the energy-momentum and angular momentum, there are two
possibilities for choosing the set of independent field
variables~\cite{kawai.toma:1991,kawai:1988,kawai.saitoh:1989a,%
kawai.saitoh:1989b}: One is to choose the set
\begin{math}
  \{\psi^{k}, A^{k}_{\hspace{0.7ex}\mu}, \phi^{A}\},
\end{math}
and the other is to choose the set
\begin{math}
  \{\psi^{k}, e^{k}_{\hspace{0.7ex}\mu}, \phi^{A}\}.
\end{math}
In the rest of this paper, we employ
\begin{math}
  \{\psi^{k}, A^{k}_{\hspace{0.7ex}\mu}, \phi^{A}\}
\end{math}
as the set of independent field variables, because this choice is
superior to the other, as shown in
Refs.~\citen{kawai.toma:1991,kawai:1988,kawai.saitoh:1989a,%
kawai.saitoh:1989b}.

From the requirement (R.i), we obtain the identities\footnote{For
  instance, $\delta \bm{L} / \delta \psi^{k}$ denotes the Euler
  derivative with respect to $\psi^{k}$.}
\begin{gather}
  \label{eq:engr_identity.i}
  \frac{\delta \bm{L}}{\delta \psi^{k}} +
  \partial_{\mu}\left(%
    \frac{\delta \bm{L}}{\delta A^{k}_{\hspace{0.7ex}\mu}}\right) +
  i\frac{\delta \bm{L}}{\delta \phi^{A}}(P_{k}\phi)^{A}
  \equiv 0, \\
  \label{eq:engr_identity.ii}
  \bm{F}_{k}^{\hspace{0.7ex}(\mu\nu)} \equiv 0, \\
  \label{eq:engr_identity.iii}
  {}^{\mathrm{tot}}\bm{T}_{k}^{\hspace{0.7ex}\mu} -
  \partial_{\nu}\bm{F}_{k}^{\hspace{0.7ex}\mu\nu} -
  \frac{\delta \bm{L}}{\delta A^{k}_{\hspace{0.7ex}\mu}}
  \equiv 0, \\
  \label{eq:engr_identity.iv}
  \partial_{\mu}{}^{\mathrm{tot}}\bm{S}_{kl}^{\hspace{1.4ex}\mu} -
  2\frac{\delta \bm{L}}{\delta \psi^{[k}}\psi_{l]} -
  2\frac{\delta \bm{L}}{\delta A^{[k}_{\hspace{1.4ex}\mu}}A_{l]\mu} -
  i\frac{\delta \bm{L}}{\delta \phi^{A}}(M_{kl}\phi)^{A} \equiv 0,
\end{gather}
where we have defined
\begin{align}
  \label{eq:lagrangian_density}
  \bm{L} &\stackrel{\mathrm{def}}{=} \sqrt{- g} L \;, \quad
  g \stackrel{\mathrm{def}}{=} \det(g_{\mu\nu}) \;, \\
  \label{eq:engr_derivative_lagrangian_diff_translation_gauge}
  \bm{F}_{k}^{\hspace{0.7ex}\mu\nu} &\stackrel{\mathrm{def}}{=}
  \frac{\partial \bm{L}}{%
    \partial A^{k}_{\hspace{0.7ex}\mu,\nu}} \;, \\
  \label{eq:engr_total_energy_momentum}
  {}^{\mathrm{tot}}\bm{T}_{k}^{\hspace{0.7ex}\mu} 
  &\stackrel{\mathrm{def}}{=}
  \frac{\partial \bm{L}}{\partial \psi^{k}_{\hspace{0.7ex},\mu}} +
  i\frac{\partial \bm{L}}{%
    \partial \phi^{A}_{\hspace{0.7ex},\mu}}(P_{k}\phi)^{A} \;, \\
  \label{eq:engr_total_spin_angular_momentum}
  {}^{\mathrm{tot}}\bm{S}_{kl}^{\hspace{1.4ex}\mu} 
  &\stackrel{\mathrm{def}}{=}
  - 2\frac{\partial \bm{L}}{%
    \partial \psi^{[k}_{\hspace{1.4ex},\mu}}\psi_{l]}
  - 2\bm{F}_{[k}^{\hspace{1.4ex}\nu\mu}A_{l]\nu}
  - i\frac{\partial \bm{L}}{%
    \partial \phi^{A}_{\hspace{0.7ex},\mu}}(M_{kl}\phi)^{A} \;.
\end{align}

If the field equations
\begin{math}
  \delta \bm{L} / \delta A^{k}_{\hspace{0.7ex}\mu} = 0
\end{math}
and 
\begin{math}
  \delta \bm{L} / \delta \phi^{A} = 0
\end{math}
are both satisfied, we have the following:
\begin{itemize}
\item The field equation
  \begin{math}
    \delta \bm{L} / \delta \psi^{k} = 0
  \end{math}
  is automatically satisfied, and thus $\psi^{k}$ is not an
  independent dynamical field variable.
\item There are two conservation laws,
  \begin{align}
    \label{eq:differential_conservation_law_emd}
    \partial_{\mu}{}^{\mathrm{tot}}\bm{T}_{k}^{\hspace{0.7ex}\mu}
    &= 0, \\
    \label{eq:differential_conservation_law_spin_amd}
    \partial_{\mu}{}^{\mathrm{tot}}\bm{S}_{kl}^{\hspace{1.4ex}\mu}
    &= 0,
  \end{align}
  which follow from
  Eqs.~(\ref{eq:engr_identity.ii})--(\ref{eq:engr_identity.iv}).
  
  The former is the differential conservation law of the dynamical
  energy-\linebreak[4] momentum, and the latter is that of the
  \lq\lq{}spin'' angular momentum.
\end{itemize}

We split the densities
\begin{math}
  {}^{\mathrm{tot}}\bm{T}_{k}^{\hspace{0.7ex}\mu}
\end{math}
and
\begin{math}
  {}^{\mathrm{tot}}\bm{S}_{kl}^{\hspace{1.4ex}\mu}
\end{math}
into gravitational and matter parts as
\begin{align}
  \label{eq:relation_split_emd}
  {}^{\mathrm{tot}}\bm{T}_{k}^{\hspace{0.7ex}\mu} &=
  {}^{G}\bm{T}_{k}^{\hspace{0.7ex}\mu} +
  {}^{M}\bm{T}_{k}^{\hspace{0.7ex}\mu} \;, \\
  \label{eq:relation_split_samd}
  {}^{\mathrm{tot}}\bm{S}_{kl}^{\hspace{1.4ex}\mu} &=
  {}^{G}\bm{S}_{kl}^{\hspace{1.4ex}\mu} +
  {}^{M}\bm{S}_{kl}^{\hspace{1.4ex}\mu} \;,
\end{align}
where we have defined
\begin{align}
  \label{eq:engr_definition_emd_gravity}
  {}^{G}\bm{T}_{k}^{\hspace{0.7ex}\mu} &\stackrel{\mathrm{def}}{=}
  \frac{\partial \bm{L}^{T}}{%
    \partial \psi^{k}_{\hspace{0.7ex},\mu}} =
  \frac{\partial \bm{L}^{T}}{%
    \partial A^{k}_{\hspace{0.7ex}\mu}} \;, \\
  \label{eq:engr_definition_emd_matter}
  {}^{M}\bm{T}_{k}^{\hspace{0.7ex}\mu} &\stackrel{\mathrm{def}}{=}
  \frac{\partial \bm{L}^{M}}{%
    \partial \psi^{k}_{\hspace{0.7ex},\mu}} +
  i\frac{\partial \bm{L}^{M}}{%
    \partial \phi^{A}_{\hspace{0.7ex},\mu}}(P_{k}\phi)^{A} =
  \frac{\partial \bm{L}^{M}}{%
    \partial A^{k}_{\hspace{0.7ex}\mu}} \;, \\
  \label{eq:engr_definition_samd_gravity}
  {}^{G}\bm{S}_{kl}^{\hspace{1.4ex}\mu} &\stackrel{\mathrm{def}}{=}
  - 2\frac{\partial \bm{L}^{T}}{%
    \partial \psi^{[k}_{\hspace{1.4ex},\mu}}\psi_{l]}
  - 2\bm{F}_{[k}^{\hspace{1.4ex}\nu\mu}A_{l]\nu} \;, \\
  \label{eq:engr_definition_samd_matter}
  {}^{M}\bm{S}_{kl}^{\hspace{1.4ex}\mu} &\stackrel{\mathrm{def}}{=}
  - 2\frac{\partial \bm{L}^{M}}{%
    \partial \psi^{[k}_{\hspace{1.4ex},\mu}}\psi_{l]}
  - i\frac{\partial \bm{L}^{M}}{%
    \partial \phi^{A}_{\hspace{0.7ex},\mu}}(M_{kl}\phi)^{A} \;,
\end{align}
with
\begin{math}
  \bm{L}^{T} \stackrel{\mathrm{def}}{=} \sqrt{- g}L^{T}
\end{math}
and
\begin{math}
  \bm{L}^{M} \stackrel{\mathrm{def}}{=} \sqrt{- g}L^{M}.
\end{math}
Here,
\begin{math}
  {}^{G}\bm{T}_{k}^{\hspace{0.7ex}\mu}
\end{math}
and
\begin{math}
  {}^{M}\bm{T}_{k}^{\hspace{0.7ex}\mu}
\end{math}
are the dynamical energy-momentum densities of the gravitational field
and the matter field, respectively, while
\begin{math}
  {}^{G}\bm{S}_{kl}^{\hspace{1.4ex}\mu}
\end{math}
and 
\begin{math}
  {}^{M}\bm{S}_{kl}^{\hspace{1.4ex}\mu}
\end{math}
are the \lq\lq{}spin'' angular momentum densities of the gravitational
field and the matter field, respectively.  The densities
\begin{math}
  {}^{G}\bm{T}_{k}^{\hspace{0.7ex}\mu},
  {}^{M}\bm{T}_{k}^{\hspace{0.7ex}\mu},
\end{math}
\begin{math}
  {}^{G}\bm{S}_{kl}^{\hspace{1.4ex}\mu}
\end{math}
and
\begin{math}
  {}^{M}\bm{S}_{kl}^{\hspace{1.4ex}\mu}
\end{math}
are all space-time vector densities~\cite{kawai:2000}.

In Ref.~\citen{kawai.toma:1991}, the integrals of the dynamical
energy-momentum and \lq\lq{}spin'' angular momentum densities over a
space-like surface $\sigma$ are examined for vierbeins with the
asymptotic behavior described below.

\noindent $\langle 1 \rangle$ The components
\begin{math}
  e^{k}_{\hspace{0.7ex}\mu}
\end{math}
of the vierbein fields possess the asymptotic property\footnote{The
  expression $O(1 / r^{n})$ with real $n$ denotes a term for which
  $r^{n} O(1 / r^{n})$ remains finite for $r \to \infty$; a term
  denoted as $O(1 / r^{n})$ could, of course, also be zero.}
\begin{equation}
  \label{eq:asymptotic_vierbein}
  e^{k}_{\hspace{0.7ex}\mu} =
  e^{(0)k}_{\hspace{2.8ex}\mu} + f^{k}_{\hspace{0.7ex}\mu} \;, \quad
  f^{k}_{\hspace{0.7ex}\mu,(m)} = O(1 / r^{1 + m}) \quad (m=0,1,2),
\end{equation}
where
\begin{math}
  f^{k}_{\hspace{0.7ex}\mu,(m)}
\end{math}
denotes the $m$th order partial derivative with respect to
$x^{\lambda}$, and the
\begin{math}
  e^{(0)k}_{\hspace{2.8ex}\mu}
\end{math}
are constant vierbeins satisfying
\begin{math}
  e^{(0)k}_{\hspace{2.8ex}\mu} \eta_{kl} 
    e^{(0)l}_{\hspace{2.8ex}\nu} = \eta_{\mu\nu}.
\end{math}

\noindent $\langle 2 \rangle$ The antisymmetric part of components
\begin{math}
  f_{\mu\nu} \stackrel{\mathrm{def}}{=} 
  e^{(0)k}_{\hspace{2.8ex}\mu} \eta_{kl}
    f^{l}_{\hspace{0.7ex}\nu}
\end{math}
satisfy
\begin{equation}
  \label{eq:asymptotic_vierbein.ii}
  f_{[\mu\nu],(m)} = O(1 / r^{1 + \alpha + m}) \quad (m=0,1),
\end{equation}
where $\alpha$ is positive but otherwise arbitrary.

It has been shown that 
\begin{align}
  \label{eq:dynamical_energy_momentum}
  M_{k} &\stackrel{\mathrm{def}}{=}
  \int_{\sigma} {}^{\mathrm{tot}}\bm{T}_{k}^{\hspace{0.7ex}\mu}
  d\sigma_{\mu} =
  e^{(0)\mu}_{\hspace{2.8ex}k} M_{\mu} \;, \\
  \label{eq:spin_angular_momentum}
  S_{kl} &\stackrel{\mathrm{def}}{=}
  \int_{\sigma} {}^{\mathrm{tot}}\bm{S}_{kl}^{\hspace{1.4ex}\mu}
  d\sigma_{\mu} =
  e^{(0)}_{\hspace{2.1ex}k\mu} e^{(0)}_{\hspace{2.1ex}l\nu}
  M^{\mu\nu} + 2\psi^{(0)}_{\hspace{2.1ex}[k} M_{l]} \;,
\end{align}
where
\begin{math}
  d\sigma_{\mu}
\end{math}
denotes the surface element on $\sigma$. In the above, we have defined
\begin{align}
  M_{\mu} &\stackrel{\mathrm{def}}{=}
  \eta_{\mu\nu} \int_{\sigma} 
  \theta^{\nu\lambda} d\sigma_{\lambda} \\
  M^{\mu\nu} &\stackrel{\mathrm{def}}{=}
  \int_{\sigma} \partial_{\rho} 
  K^{\mu\nu\lambda\rho} d\sigma_{\lambda} =
  \int_{\sigma}(
    x^{\mu}\theta^{\nu\lambda} - x^{\nu}\theta^{\mu\lambda})
  d\sigma_{\lambda} \;,
\end{align}
with
\begin{align}
  \label{eq:symmetric_energy_momentum_density.LL}
  \theta^{\nu\lambda} &\stackrel{\mathrm{def}}{=}
  \frac{1}{\kappa} \partial_{\rho}\partial_{\sigma} \{%
    (- g)g^{\nu[\lambda}g^{\rho]\sigma} \}, \\
  K^{\mu\nu\lambda\rho} &\stackrel{\mathrm{def}}{=}
  \frac{1}{\kappa} \left(
    x^{\mu} \partial_{\sigma}\{(- g)g^{\nu[\lambda}g^{\rho]\sigma}\} -
    x^{\nu} \partial_{\sigma}\{(- g)g^{\mu[\lambda}g^{\rho]\sigma}\} +
    (- g)g^{\mu[\lambda}g^{\rho]\nu} - 
    \eta^{\mu[\lambda}\eta^{\rho]\nu} \right). \nonumber \\
\end{align}
This expression of 
\begin{math}
  \theta^{\nu\lambda}
\end{math}
is the same as that of the symmetric energy-momentum density proposed
by Landau and Lifshitz~\cite{landau.lifshitz}.  Equation
(\ref{eq:spin_angular_momentum}) has been obtained by choosing the
asymptotic form of the Higgs-type field $\psi^{k}$ as
\begin{subequations}
  \label{eq:higgs_type_field_asymptotic_form.i}
  \begin{align}
    \label{eq:higgs_type_field_asymptotic_form.ia}
    \psi^{k} &=
    e^{(0)k}_{\hspace{2.8ex}\mu} x^{\mu} +
    \psi^{(0)k} + O\left(%
      \frac{1}{r^{\beta}}\right), \\
    \label{eq:higgs_type_field_asymptotic_form.ib}
    \psi^{k}_{\hspace{0.7ex},\mu} &=
    e^{(0)k}_{\hspace{2.8ex}\mu} + O\left(%
      \frac{1}{r^{1 + \beta}}\right), \quad (\beta > 0) \\
    \label{eq:higgs_type_field_asymptotic_form.ic}
    \psi^{k}_{\hspace{0.7ex},\mu\nu} &=
    O\left(\frac{1}{r^{2}}\right),
  \end{align}
\end{subequations}
with
\begin{math}
  \psi^{(0)k}
\end{math}
and $\beta$ constant, whereas Eq.~(\ref{eq:dynamical_energy_momentum})
has been obtained without imposing the conditions
(\ref{eq:higgs_type_field_asymptotic_form.i}).

From the requirement (R.ii), we obtain the identity
\begin{equation}
  \label{eq:engr_identity.v}
  \widetilde{\bm{T}}_{\mu}^{\hspace{0.7ex}\nu} -
  \partial_{\lambda}\bm{\Psi}_{\mu}^{\hspace{0.7ex}\nu\lambda} -
  \frac{\delta \bm{L}}{\delta A^{k}_{\hspace{0.7ex}\nu}}
    A^{k}_{\hspace{0.7ex}\mu} \equiv 0,
\end{equation}
with
\begin{align}
  \label{eq:engr_canonical_emd}
  \widetilde{\bm{T}}_{\mu}^{\hspace{0.7ex}\nu}
  &\stackrel{\mathrm{def}}{=}
  \delta_{\mu}^{\hspace{0.7ex}\nu} \bm{L} -
  \bm{F}_{k}^{\hspace{0.7ex}\lambda\nu}
    A^{k}_{\hspace{0.7ex}\lambda,\mu} -
  \frac{\partial \bm{L}}{\partial \phi^{A}_{\hspace{0.7ex},\nu}}
    \phi^{A}_{\hspace{0.7ex},\mu} -
  \frac{\partial \bm{L}}{\partial \psi^{k}_{\hspace{0.7ex},\nu}}
    \psi^{k}_{\hspace{0.7ex},\mu} \;, \\
  \label{eq:definition_Phi}
  \bm{\Psi}_{\mu}^{\hspace{0.7ex}\nu\lambda}
  &\stackrel{\mathrm{def}}{=}
  \bm{F}_{k}^{\hspace{0.7ex}\nu\lambda} A^{k}_{\hspace{0.7ex}\mu} =
  - \bm{\Psi}_{\mu}^{\hspace{0.7ex}\lambda\nu} \;.
\end{align}
The identity (\ref{eq:engr_identity.v}) leads to
\begin{align}
  \label{eq:differential_conservation_law_canonical_emd}
  \partial_{\nu} \widetilde{\bm{T}}_{\mu}^{\hspace{0.7ex}\nu} &= 0, \\
  \label{eq:differential_conservation_law_extended_oamd}
  \partial_{\lambda}
    \widetilde{\bm{M}}_{\mu}^{\hspace{0.7ex}\nu\lambda} &= 0
\end{align}
when
\begin{math}
  \delta \bm{L} / \delta A^{k}_{\hspace{0.7ex}\nu} = 0,
\end{math}
where we have defined
\begin{equation}
  \label{eq:engr_extended_orbital_amd}
  \widetilde{\bm{M}}_{\mu}^{\hspace{0.7ex}\nu\lambda}
  \stackrel{\mathrm{def}}{=}
  2(\bm{\Psi}_{\mu}^{\hspace{0.7ex}\nu\lambda} -
  x^{\nu}\widetilde{\bm{T}}_{\mu}^{\hspace{0.7ex}\lambda}).
\end{equation}
Equations (\ref{eq:differential_conservation_law_canonical_emd}) and
(\ref{eq:differential_conservation_law_extended_oamd}) are the
differential conservation laws of the canonical energy-momentum and
the \lq\lq{}extended orbital angular momentum'' defined by
\begin{equation}
  M^{c}_{\hspace{0.7ex}\mu} \stackrel{\mathrm{def}}{=}
  \int_{\sigma} \widetilde{\bm{T}}_{\mu}^{\hspace{0.7ex}\nu}
  d\sigma_{\nu} \;, \quad
  L_{\mu}^{\hspace{0.7ex}\nu} \stackrel{\mathrm{def}}{=}
  \int_{\sigma} \widetilde{\bm{M}}_{\mu}^{\hspace{0.7ex}\nu\lambda}
  d\sigma_{\lambda} \;,
\end{equation}
respectively~\cite{kawai.toma:1991}.  The canonical energy-momentum
and the \lq\lq{}extended orbital angular momentum'' are the generators
of general affine \emph{coordinate} transformations. The antisymmetric
part
\begin{math}
  L_{[\mu\nu]} \stackrel{\mathrm{def}}{=}
  L_{[\mu}^{\hspace{1.0ex}\lambda} \eta_{\lambda\nu]}
\end{math}
is the orbital angular momentum\footnote{This $L_{[\mu\nu]}$ should
  not be confused with the orbital part of the \lq\lq{}spin'' angular
  momentum $S_{kl}$.} and is the generator of \emph{coordinate}
Lorentz transformation~\cite{kawai.toma:1991}.

We split the canonical energy-momentum density into gravitational
and matter parts as
\begin{equation}
  \widetilde{\bm{T}}_{\mu}^{\hspace{0.7ex}\nu} =
  {}^{G}\widetilde{\bm{T}}_{\mu}^{\hspace{0.7ex}\nu} +
  {}^{M}\bm{T}_{\mu}^{\hspace{0.7ex}\nu} \;,
\end{equation}
where we have defined
\begin{align}
  \label{eq:canonical_emd_gravity}
  {}^{G}\widetilde{\bm{T}}_{\mu}^{\hspace{0.7ex}\nu}
  &\stackrel{\mathrm{def}}{=}
  \delta_{\mu}^{\hspace{0.7ex}\nu}\bm{L}^{T} -
  \bm{F}_{k}^{\hspace{0.7ex}\lambda\nu}
    A^{k}_{\hspace{0.7ex}\lambda,\mu} -
  \frac{\partial \bm{L}^{T}}{\partial \psi^{k}_{\hspace{0.7ex},\nu}}
    \psi^{k}_{\hspace{0.7ex},\mu} \;, \\
  \label{eq:canonical_emd_matter}
  {}^{M}\bm{T}_{\mu}^{\hspace{0.7ex}\nu}
  &\stackrel{\mathrm{def}}{=}
  \delta_{\mu}^{\hspace{0.7ex}\nu}\bm{L}^{M} -
  \frac{\partial \bm{L}^{M}}{\partial \psi^{k}_{\hspace{0.7ex},\nu}}
    \psi^{k}_{\hspace{0.7ex},\mu} -
  \frac{\partial \bm{L}^{M}}{\partial \phi^{A}_{\hspace{0.7ex},\nu}}
    \phi^{A}_{\hspace{0.7ex},\mu} \;.
\end{align}
The density
\begin{math}
  {}^{G}\widetilde{\bm{T}}_{\mu}^{\hspace{0.7ex}\nu}
\end{math}
does not transform as a tensor density under general coordinate
transformations, while
\begin{math}
  {}^{M}\bm{T}_{\mu}^{\hspace{0.7ex}\nu}
\end{math}
does transform as a tensor density.\cite{kawai:2000}

As is described in Ref.~\citen{kawai.toma:1991}, the generators
\begin{math}
  M^{c}_{\hspace{0.7ex}\mu}
\end{math}
and
\begin{math}
  L_{\mu}^{\hspace{0.7ex}\nu}
\end{math}
vanish for vierbeins with the asymptotic forms satisfying
Eqs.~(\ref{eq:asymptotic_vierbein}) and
(\ref{eq:asymptotic_vierbein.ii}) when the condition
(\ref{eq:higgs_type_field_asymptotic_form.i}) is satisfied.

The field equation
\begin{math}
  \delta \bm{L} / \delta A^{k}_{\hspace{0.7ex}\mu} = 0
\end{math}
has the expression
\begin{equation}
  \label{eq:field_eq_translation_gauge}
  - 2\nabla_{\lambda}F^{\mu\nu\lambda}
  + 2 v_{\lambda}F^{\mu\nu\lambda}
  + 2 H^{\mu\nu}
  - g^{\mu\nu}L^{T} = T^{\mu\nu} \;,
\end{equation}
where we have defined
\begin{align}
  \label{eq:covariant_derivative_F}
  \nabla_{\lambda}F^{\mu\nu\lambda} &\stackrel{\mathrm{def}}{=}
  \partial_{\lambda}F^{\mu\nu\lambda} +
  \Gamma^{\mu}_{\hspace{0.7ex}\sigma\lambda}
    F^{\sigma\nu\lambda} +
  \Gamma^{\nu}_{\hspace{0.7ex}\sigma\lambda}
    F^{\mu\sigma\lambda} +
  \Gamma^{\lambda}_{\hspace{0.7ex}\sigma\lambda}
    F^{\mu\nu\sigma} \;, \\
  \label{eq:explicit_form_F}
  F^{\mu\nu\lambda} &\stackrel{\mathrm{def}}{=}
  c_{1}(t^{\mu\nu\lambda} - t^{\mu\lambda\nu}) +
  c_{2}(g^{\mu\nu}v^{\lambda} - g^{\mu\lambda}v^{\nu}) -
  \frac{1}{3}c_{3} \epsilon^{\mu\nu\lambda\rho}a_{\rho} \;, \\
  \label{eq:explicit_form_H}
  H^{\mu\nu} &\stackrel{\mathrm{def}}{=}
  T^{\rho\sigma\mu}F_{\rho\sigma}^{\hspace{1.4ex}\nu} -
  \frac{1}{2}T^{\nu\rho\sigma}F^{\mu}_{\hspace{0.7ex}\rho\sigma} \;.
\end{align}
Also, 
\begin{math}
  T^{\mu\nu}
\end{math}
is the energy-momentum density of the gravitational source defined by
\begin{equation}
  \label{eq:gravitational_source}
  \sqrt{- g} T^{\mu\nu} \stackrel{\mathrm{def}}{=}
  \eta^{kl}e^{\mu}_{\hspace{0.7ex}l} \frac{\delta \bm{L}^{M}}{%
    \delta A^{k}_{\hspace{0.7ex}\nu}} \;.
\end{equation}

\section{Weak-field approximation}
\label{sec:weak_field_approximation}

We now consider weak field situations in which the vierbein fields
\begin{math}
  e^{k}_{\hspace{0.7ex}\mu}
\end{math}
take the form
\begin{equation}
  \label{eq:weak_vierbein}
  e^{k}_{\hspace{0.7ex}\mu} =
  e^{(0)k}_{\hspace{2.8ex}\mu} + f^{k}_{\hspace{0.7ex}\mu} \;, \quad
  |f^{k}_{\hspace{0.7ex}\mu}| \ll 1,
\end{equation}
where the
\begin{math}
  e^{(0)k}_{\hspace{2.8ex}\mu}
\end{math}
are constant vierbeins satisfying
\begin{math}
  e^{(0)k}_{\hspace{2.8ex}\mu} \eta_{kl} e^{(0)l}_{\hspace{2.8ex}\nu}
  = \eta_{\mu\nu}.
\end{math}
The components of the metric and torsion tensors are given by, up to
terms linear in
\begin{math}
  f^{k}_{\hspace{0.7ex}\mu},
\end{math}
\begin{align}
  \label{eq:weak_metric}
  g_{\mu\nu} &= \eta_{\mu\nu} + 2f_{(\mu\nu)} \;, \\
  \label{eq:weak_torsion}
  T_{\lambda\mu\nu} &=
  \partial_{\mu}f_{\lambda\nu} - \partial_{\nu}f_{\lambda\mu} \;,
\end{align}
where we have defined
\begin{math}
  f_{\mu\nu} \stackrel{\mathrm{def}}{=}
  e^{(0)k}_{\hspace{2.8ex}\mu} \eta_{kl} f^{l}_{\hspace{0.7ex}\nu}
\end{math}.\footnote{For $f^{k}_{\hspace{0.7ex}\mu}$, we use the
  convention that both the Greek and Latin indices of
  $f^{k}_{\hspace{0.7ex}\mu}$ are raised or lowered with the Minkowski
  metric, and that they are converted into one another with
  $e^{(0)k}_{\hspace{2.8ex}\mu}$ or $e^{(0)\mu}_{\hspace{2.8ex}k}$,
  where $(e^{(0)\mu}_{\hspace{2.8ex}k}) \stackrel{\mathrm{def}}{=}
  (e^{(0)k}_{\hspace{2.8ex}\mu})^{- 1}$. Thus, $f^{\mu\nu}$, for
  example, represents $\eta^{\nu\lambda} e^{(0)\mu}_{\hspace{2.8ex}k}
  f^{k}_{\hspace{0.7ex}\lambda} (\neq g^{\nu\lambda}
  e^{\mu}_{\hspace{0.7ex}k} f^{k}_{\hspace{0.7ex}\lambda})$.}

In the weak-field approximation, the symmetric and antisymmetric parts
of the field equation (\ref{eq:field_eq_translation_gauge}) take the
form
\begin{multline}
  \label{eq:weak_symmetric_field_eq}
  3c_{1} \left\{ \square \bar{f}_{(\mu\nu)} -
    \partial^{\lambda}(\partial_{\mu} \bar{f}_{(\nu\lambda)} +
    \partial_{\nu} \bar{f}_{(\mu\lambda)}) +
    \eta_{\mu\nu} \partial_{\rho}\partial_{\sigma} 
    \bar{f}^{(\rho\sigma)} \right\} \\
  \hspace*{-10ex}+ (c_{1} + c_{2}) \left\{
    - \eta_{\mu\nu} \square \bar{f} - 
    2\eta_{\mu\nu} \partial_{\rho}\partial_{\sigma}
    \bar{f}^{(\rho\sigma)} +
    \partial_{\mu}\partial_{\nu} \bar{f} \right. \\
  \qquad + \left.
    \partial^{\lambda}(\partial_{\mu} \bar{f}_{(\nu\lambda)} +
    \partial_{\nu} \bar{f}_{(\mu\lambda)}) -
    \partial^{\lambda}(\partial_{\mu} f_{[\nu\lambda]} +
    \partial_{\nu} f_{[\mu\lambda]}) \right\} = T_{(\mu\nu)} \;,
\end{multline}
\begin{multline}
  \label{eq:weak_antisymmetric_field_eq}
  \left(c_{1} - \frac{4}{9}c_{3}\right) \left\{
    \square f_{[\mu\nu]} +
    \partial^{\lambda}(\partial_{\mu} f_{[\nu\lambda]} -
    \partial_{\nu} f_{[\mu\lambda]}) \right\} \\
  + (c_{1} + c_{2}) \left\{
    \partial^{\lambda}(\partial_{\mu} \bar{f}_{(\nu\lambda)} -
    \partial_{\nu} \bar{f}_{(\mu\lambda)}) -
    \partial^{\lambda}(\partial_{\mu} f_{[\nu\lambda]} -
    \partial_{\nu} f_{[\mu\lambda]}) \right\} = T_{[\mu\nu]} \;,
\end{multline}
with
\begin{math}
  \square \stackrel{\mathrm{def}}{=} \partial^{\mu} \partial_{\mu}
\end{math}.
Here, we have introduced
\begin{align}
  \label{eq:definition_bar_f}
  \bar{f}_{(\mu\nu)} &\stackrel{\mathrm{def}}{=}
  f_{(\mu\nu)} - \frac{1}{2}\eta_{\mu\nu}f \;, \quad
  f \stackrel{\mathrm{def}}{=} \eta^{\mu\nu}f_{(\mu\nu)} \;, \\
  \label{eq:definition_bar_f_trace}
  \bar{f} &\stackrel{\mathrm{def}}{=}
  \eta^{\mu\nu}\bar{f}_{(\mu\nu)} \;.
\end{align}

We consider the energy-momentum density of the source
\begin{math}
  T_{\mu\nu}
\end{math}
to lowest order in 
\begin{math}
  f^{k}_{\hspace{0.7ex}\mu}.
\end{math}
Therefore it is independent of
\begin{math}
  f^{k}_{\hspace{0.7ex}\mu}
\end{math}
and satisfies the ordinary conservation law in special relativity,
\begin{equation}
  \label{eq:ordinary_conservation_law}
  \partial^{\nu}T_{\mu\nu} = 0.
\end{equation}

Let us consider the transformations
\begin{align}
  \label{eq:general_coordinate_transformation.i}
  f'_{(\mu\nu)} &=
  f_{(\mu\nu)} - \partial_{\mu}\varepsilon_{\nu} -
  \partial_{\nu}\varepsilon_{\mu} \;, \\
  \label{eq:general_coordinate_transformation.ii}
  f'_{[\mu\nu]} &= f_{[\mu\nu]} + \partial_{\mu}\chi_{\nu} - 
    \partial_{\nu}\chi_{\mu} \;,
\end{align}
where
\begin{math}
  \varepsilon_{\mu}
\end{math}
and
\begin{math}
  \chi_{\mu}
\end{math}
are arbitrary small functions.  Since
Eqs.~(\ref{eq:weak_symmetric_field_eq}) and
(\ref{eq:weak_antisymmetric_field_eq}) are invariant under the
transformations (\ref{eq:general_coordinate_transformation.i}) and
(\ref{eq:general_coordinate_transformation.ii}) with
\begin{math}
  \varepsilon_{\mu} = \chi_{\mu},
\end{math}
we can impose the harmonic coordinate condition
\begin{equation}
  \label{eq:harmonic_coordinate_conditions}
  \partial_{\nu} \bar{f}^{(\mu\nu)} = 0.
\end{equation}
The Lagrangian $L^{T}$ with the parameters $c_{1}$ and $c_{2}$
satisfying
\begin{equation}
  \label{eq:parameters_conditions}
  c_{1} = - c_{2} = - \frac{1}{3\kappa}
\end{equation}
compares quite favorably with
experiment~\cite{hayashi.shirafuji:1979}.  We therefore assume
(\ref{eq:parameters_conditions}) to hold henceforth.  Under the
conditions (\ref{eq:harmonic_coordinate_conditions}) and
(\ref{eq:parameters_conditions}),
Eq.~(\ref{eq:weak_antisymmetric_field_eq}) reduces to
\begin{equation}
  \label{eq:weak_antisymmetric_field_eq.ii}
  \left(c_{1} - \frac{4}{9}c_{3}\right) \left\{%
    \square f_{[\mu\nu]} +
    \partial^{\lambda}(\partial_{\mu} f_{[\nu\lambda]} -
      \partial_{\nu} f_{[\mu\lambda]})
  \right\} = T_{[\mu\nu]} \;,
\end{equation}
which is still invariant under the transformation
(\ref{eq:general_coordinate_transformation.ii})~\cite{%
hayashi.shirafuji:1979}. Thus, we can impose the condition
\begin{equation}
  \label{eq:harmonic_coordinate_conditions.ii}
  \partial_{\nu}f^{[\mu\nu]} = 0.
\end{equation}
Finally, under the conditions
(\ref{eq:harmonic_coordinate_conditions}),
(\ref{eq:parameters_conditions}) and
(\ref{eq:harmonic_coordinate_conditions.ii}), the field equations of
\begin{math}
  \bar{f}_{(\mu\nu)}
\end{math}
and 
\begin{math}
  f_{[\mu\nu]}
\end{math}
become
\begin{align}
  \label{eq:wave_equation_symmetric}
  \square \bar{f}_{(\mu\nu)} &= - \kappa T_{(\mu\nu)} \;, \\
  \label{eq:wave_equation_antisymmetric}
  \square f_{[\mu\nu]} &= - \lambda T_{[\mu\nu]} \;,
\end{align}
where we have defined 
\begin{math}
  1 / \lambda \stackrel{\mathrm{def}}{=}
  - c_{1} + (4/9)c_{3} \neq 0.
\end{math}
From Eqs.~(\ref{eq:harmonic_coordinate_conditions}) and
(\ref{eq:wave_equation_symmetric}), we find that the symmetric part of
\begin{math}
  T_{\mu\nu}
\end{math}
satisfies the conservation law
\begin{equation}
  \label{eq:ordinary_conservation_law_symmetric}
  \partial^{\nu} T_{(\mu\nu)} = 0,
\end{equation}
and the antisymmetric part of
\begin{math}
  T_{\mu\nu}
\end{math}
satisfies
\begin{equation}
  \label{eq:ordinary_conservation_law_antisymmetric}
  \partial^{\nu} T_{[\mu\nu]} = 0,
\end{equation}
which follows from Eqs.~(\ref{eq:harmonic_coordinate_conditions.ii})
and
(\ref{eq:wave_equation_antisymmetric})~\cite{hayashi.shirafuji:1979}.

Let us consider the plane wave solutions of
Eqs.~(\ref{eq:wave_equation_symmetric}) and
(\ref{eq:wave_equation_antisymmetric}) with
\begin{math}
  T_{\mu\nu} \equiv 0,
\end{math}
\begin{align}
  \label{eq:plane_wave_solution_symmetric}
  \bar{f}_{(\mu\nu)}(\bm{x}, x^{0}) &=
  {\mathcal{U}}_{\mu\nu} e^{ik \cdot x} +
  \bar{\mathcal{U}}_{\mu\nu} e^{- ik \cdot x} \;, \\
  \label{eq:plane_wave_solution_antisymmetric}
  f_{[\mu\nu]}(\bm{x}, x^{0}) &=
  {\mathcal{V}}_{\mu\nu}e^{ik \cdot x} +
  \bar{\mathcal{V}}_{\mu\nu}e^{- ik \cdot x} \;,
\end{align}
where
\begin{math}
  k \cdot x \stackrel{\mathrm{def}}{=}
  k_{\mu} x^{\mu}.
\end{math}
Here,
\begin{math}
  \mathcal{U}_{\mu\nu}
\end{math}
and
\begin{math}
  \mathcal{V}_{\mu\nu}
\end{math}
are constant amplitudes,
\begin{math}
  \bar{\mathcal{U}}
\end{math}
and
\begin{math}
  \bar{\mathcal{V}}
\end{math}
are their complex conjugates, and $k_{\mu}$ is a constant wave vector,
which satisfy the relations
\begin{equation}
  \label{eq:null_and_orthogonal_conditions}
  k_{\mu}k^{\mu} = 0, \quad
  {\mathcal{U}}_{\mu\nu}k^{\nu} = 0, \quad
  {\mathcal{V}}_{\mu\nu} k^{\nu} = 0.
\end{equation}
Following the prescription given in Section 35.4 of
Ref.~\citen{misner.thorne.wheeler}, we impose the
transverse-traceless gauge condition
\begin{equation}
  \label{eq:TT_gauge_conditions}
  {\mathcal{U}}_{\mu\nu}\zeta^{\nu} = 0, \quad
  {\mathcal{U}}^{\mu}_{~\mu} = 0, \quad
  {\mathcal{V}}_{\mu\nu} \zeta^{\nu} = 0,
\end{equation}
where $\zeta^{\mu}$ is a constant time-like vector. We see that the
number of physically significant components of
\begin{math}
  \bar{f}_{(\mu\nu)}
\end{math}
is two, while that of
\begin{math}
  f_{[\mu\nu]}
\end{math}
is one.

We next calculate the energy-momentum of the plane waves given by
Eqs.~(\ref{eq:plane_wave_solution_symmetric}) and
(\ref{eq:plane_wave_solution_antisymmetric}).  The dynamical
energy-momentum density
\begin{math}
  {}^{G}\bm{T}_{l}^{\hspace{0.7ex}\mu}
\end{math}
of gravitational field has, to lowest order in
\begin{math}
  f_{\mu\nu},
\end{math}
the expression
\begin{align}
  \label{eq:weak_emd_gravity_ENGR}
  2\kappa \; {}^{G}\bm{T}_{l}^{~\mu} &=
  e^{(0)\mu}_{\hspace{2.8ex}l} \biggl[
    - \partial^{\sigma}\bar{f}^{(\lambda\rho)}
      \partial_{\sigma}\bar{f}_{(\lambda\rho)}
    + \partial^{\sigma}\bar{f}^{(\lambda\rho)}
      \partial_{\rho}\bar{f}_{(\lambda\sigma)}
    + \frac{1}{2} \partial^{\sigma}\bar{f}
      \partial_{\sigma}\bar{f} 
    + 2 \partial^{\sigma}\bar{f}^{(\lambda\rho)}
      \partial_{\rho}f_{[\lambda\sigma]} \notag \\
  & \quad
    + \partial^{\lambda}f^{[\sigma\rho]}
      \partial_{\rho}f_{[\lambda\sigma]}
    - \frac{\kappa}{\lambda} \Bigl(
        \partial^{\sigma}f^{[\lambda\rho]}
        \partial_{\sigma}f_{[\lambda\rho]}
      + 2\partial^{\lambda}f^{[\sigma\rho]}
        \partial_{\rho}f_{[\lambda\sigma]} \Bigr)
    \biggr] \notag \\
  &
  - 2e^{(0)\nu}_{\hspace{2.8ex}l} \biggl[
    - \partial^{\mu}\bar{f}^{(\rho\sigma)}
      \partial_{\nu}\bar{f}_{(\rho\sigma)}
    + \frac{1}{2} \partial^{\mu}\bar{f}\partial_{\nu}\bar{f} 
    - \partial^{\sigma}\bar{f}^{(\rho\mu)}
      \partial_{\sigma}\bar{f}_{(\rho\nu)}
    + \partial^{\sigma}\bar{f}^{(\rho\mu)}
      \partial_{\nu}\bar{f}_{(\rho\sigma)} \notag \\
  & \quad
    + \partial^{\mu}\bar{f}^{(\rho\sigma)}
      \partial_{\sigma}\bar{f}_{(\rho\nu)}
    + \frac{1}{2}\partial^{\sigma}\bar{f}^{(\rho\mu)}
      \eta_{\rho\nu}\partial_{\sigma}\bar{f}
    - \frac{1}{2}\partial^{\mu}\bar{f}_{(\nu\sigma)}
      \partial^{\sigma}\bar{f}
    - \partial^{\sigma}\bar{f}^{(\rho\mu)}
      \partial_{\sigma}f_{[\rho\nu]} \notag \\
  & \quad 
    + \partial^{\sigma}\bar{f}^{(\rho\mu)}
      \partial_{\nu}f_{[\rho\sigma]}
    + \partial^{\mu}\bar{f}^{(\rho\sigma)}
      \partial_{\sigma}f_{[\rho\nu]}
    + \partial^{\sigma}f^{[\rho\mu]}
      \partial_{\nu}\bar{f}_{(\rho\sigma)}
    - \partial^{\rho}f^{[\sigma\mu]}
      \partial_{\sigma}\bar{f}_{(\rho\nu)} \notag \\
  & \quad
    + \frac{1}{2} \partial_{\nu}f^{[\sigma\mu]}
      \partial_{\sigma}\bar{f}
    - \partial^{\sigma}f^{[\rho\mu]}
      \partial_{\nu}f_{[\rho\sigma]}
    - \partial^{\rho}f^{[\sigma\mu]}
      \partial_{\sigma}f_{[\rho\nu]} \notag \\
  & \quad
  - \frac{\kappa}{\lambda} \Bigl(
      \partial^{\sigma}f^{[\rho\mu]}
      \partial_{\sigma}\bar{f}_{(\rho\nu)}
    - \partial^{\mu}f^{[\rho\sigma]}
      \partial_{\sigma}\bar{f}_{(\rho\nu)}
    - \partial^{\rho}f^{[\sigma\mu]}
      \partial_{\sigma}\bar{f}_{(\rho\nu)} \notag \\
  & \qquad
    - \frac{1}{2}\partial^{\sigma}f^{[\rho\mu]}
      \eta_{\rho\nu}\partial_{\sigma}\bar{f}
    + \frac{1}{2}\partial^{\mu}f_{[\nu\sigma]}
      \partial^{\sigma}\bar{f}
    + \frac{1}{2}\partial_{\nu}f^{[\sigma\mu]}
      \partial_{\sigma}\bar{f}
    + \partial^{\sigma}f^{[\rho\mu]}
    \partial_{\sigma}f_{[\rho\nu]} \notag \\
  & \qquad
    - 2 \partial^{\sigma}f^{[\rho\mu]}
    \partial_{\nu}f_{[\rho\sigma]}
    - \partial^{\mu}f^{[\rho\sigma]}
    \partial_{\sigma}f_{[\rho\nu]}
    + \partial^{\mu}f^{[\rho\sigma]}
    \partial_{\nu}f_{[\rho\sigma]} \notag \\
  & \qquad
    - \partial^{\rho}f^{[\sigma\mu]}
    \partial_{\sigma}f_{[\rho\nu]} \Bigr)\biggr].
\end{align}
By using Eqs.~(\ref{eq:plane_wave_solution_symmetric})--(%
\ref{eq:TT_gauge_conditions}),
\begin{math}
  {}^{G}\bm{T}_{l}^{\hspace{0.7ex}\mu}
\end{math}
is found to be given by
\begin{align}
  \label{eq:weak_emd_TT_gauge}
  {}^{G}\bm{T}_{l}^{\hspace{0.7ex}\mu} &=
  - 2e^{(0)\nu}_{\hspace{2.8ex}l} \left[
    \frac{k^{\mu}k_{\nu}}{2\kappa}\left(
      {\mathcal{U}}^{\rho\sigma}
        {\mathcal{U}}_{\rho\sigma}e^{2ik \cdot x} -
      2 {\mathcal{U}}^{\rho\sigma}\bar{\mathcal{U}}_{\rho\sigma} +
      \bar{\mathcal{U}}^{\rho\sigma}\bar{\mathcal{U}}_{\rho\sigma}
        e^{- 2ik \cdot x} \right)
  \right. \notag \\ 
  & \qquad \left. +
    \frac{k^{\mu}k_{\nu}}{2\lambda}\left(
      {\mathcal{V}}^{\rho\sigma}
        {\mathcal{V}}_{\rho\sigma}e^{2ik \cdot x} -
      2 {\mathcal{V}}^{\rho\sigma}\bar{\mathcal{V}}_{\rho\sigma} +
      \bar{\mathcal{V}}^{\rho\sigma}\bar{\mathcal{V}}_{\rho\sigma}
        e^{- 2ik \cdot x} \right)
  \right].
\end{align}
Taking the average of
\begin{math}
  {}^{G}\bm{T}_{l}^{\hspace{0.7ex}\mu}
\end{math}
over a space-time region much larger than 
\begin{math}
  |\bm{k}|^{- 1},
\end{math}
we obtain
\begin{equation}
  \label{eq:wave_length_average}
  \langle {}^{G}\bm{T}_{l}^{\hspace{0.7ex}\mu} \rangle =
  e^{(0)\nu}_{\hspace{2.8ex}l} \left[
    \frac{c^{4}k^{\mu}k_{\nu}}{2\pi G} \left(
      |{\mathcal{U}}_{11}|^{2} + |{\mathcal{U}}_{12}|^{2} \right) +
    4 \frac{k^{\mu}k_{\nu}}{\lambda} |{\mathcal{V}}_{12}|^{2} 
  \right],
\end{equation}
where we have chosen the direction of the space components $\bm{k}$
of the four vector $k^{\mu}$ as the third axis. The term in the square
brackets of the right-hand side (r.h.s.) of
Eq.~(\ref{eq:wave_length_average}) is identical to the corresponding
term of the canonical energy-momentum density
\begin{math}
  \tilde{\bm{t}}_{\nu}^{\hspace{0.7ex}\mu}
\end{math}
defined by Eq.~(\ref{eq:canonical_emd.GR}) in GR if the antisymmetric
part
\begin{math}
  {\mathcal{V}}_{12}
\end{math}
is vanishing.\footnote{See, for instance, Section 10.3 of
  Ref.~\citen{weinberg}}

\section{Quadrupole radiation from point masses}
\label{sec:quadrupole}

The retarded solutions of Eqs.~(\ref{eq:wave_equation_symmetric}) and
(\ref{eq:wave_equation_antisymmetric}) are given by
\begin{align}
  \label{eq:retarded_solution_symmetric}
  \bar{f}_{(\mu\nu)}(\bm{x},x^{0}) &=
  \frac{\kappa}{4\pi}\int d^{3}x'
  \frac{T_{(\mu\nu)}(\bm{x}',
    x^{0} - |\bm{x} - \bm{x}'|)}{%
    |\bm{x} - \bm{x}'|} \;, \\
  \label{eq:retarded_solution_antisymmetric}
  f_{[\mu\nu]}(\bm{x},x^{0}) &=
  \frac{\lambda}{4\pi}\int d^{3}x'
  \frac{T_{[\mu\nu]}(\bm{x}',
    x^{0} - |\bm{x} - \bm{x}'|)}{%
    |\bm{x} - \bm{x}'|} \;,
\end{align}
respectively. We consider a system of the Newtonian point masses
\begin{math}
  \{m_{a}, \xi^{\mu}_{a}\} \; (a = 1,2, \ldots, N)
\end{math}
as a gravitational wave source, where $m_{a}$ and $\xi^{\mu}_{a}$
denote the mass and coordinate of the $a$th point mass, respectively.
For this case, the antisymmetric part of the energy-momentum density
becomes
\begin{equation}
  \label{eq:anti_symmetric_part_emd_point_masses}
  T_{[\mu\nu]}(\bm{x}, x^{0}) = 0.
\end{equation}
Thus,
\begin{math}
  f_{[\mu\nu]}
\end{math}
vanishes, i.e., 
\begin{math}
  f_{[\mu\nu]} = 0.
\end{math}
Applying the retarded expansion to
Eq.~(\ref{eq:retarded_solution_symmetric}) and using the conservation
law (\ref{eq:ordinary_conservation_law_symmetric}), we obtain the
quadrupole radiation formula in the rest frame of the system,
\begin{subequations}
  \label{eq:general_quadrupole_formula}
  \begin{align}
    \label{eq:general_quadrupole_formula_ss}
    \bar{f}_{(\alpha\beta)}(\bm{x}, x^{0}) &=
    \frac{\kappa}{8\pi r} \partial_{0}\partial_{0}
      \int d^{3}x' x^{\prime\alpha}x^{\prime\beta}
      T_{(00)}(\bm{x}', x^{0} - r), \\
    \bar{f}_{(0\alpha)}(\bm{x}, x^{0}) &= - 
    \frac{x^{\beta}}{r} 
      \bar{f}_{(\alpha\beta)}(\bm{x}', x^{0} - r), \\
    \bar{f}_{(00)}(\bm{x}, x^{0}) &=
    \frac{\kappa}{4\pi r}\sum_{a = 1}^{N} m_{a}c^{2} +
    \frac{x^{\alpha}x^{\beta}}{r^{2}}
      \bar{f}_{(\alpha\beta)}(\bm{x}', x^{0} - r),
  \end{align}
\end{subequations}
with
\begin{math}
  r = (x^{\alpha}x^{\alpha})^{\frac{1}{2}}
\end{math}.\footnote{In our convention, letters from the beginning of
  the Greek alphabet, $\alpha$, $\beta$, $\gamma$, $\cdots$, and those
  from the beginning of the Latin alphabet, $a$, $b$, $c$, $\cdots$,
  take the values 1, 2, and 3, unless otherwise stated. Also, here we
  have used the usual summation convention for repeated indices. }\
Since we regard each velocity of the mass
points to be much smaller than the velocity of light, we can use the
quantity
\begin{equation}
  \label{eq:energy_density_of_point_masses}
  T_{(00)}(\bm{x}, x^{0}) =
  \sum_{a=1}^{N} m_{a}c^{2} 
    \delta^{(3)}(\bm{x} - \bm{\xi}_{a}(t))
\end{equation}
as the energy density of the system, with
\begin{math}
  t \stackrel{\mathrm{def}}{=} x^{0}/c.
\end{math}
Substituting Eq.~(\ref{eq:energy_density_of_point_masses}) into
Eq.~(\ref{eq:general_quadrupole_formula}), we obtain
\begin{subequations}
  \label{eq:quadrupole_formula}
  \begin{align}
    \bar{f}_{(\alpha\beta)} &=
    \frac{\kappa}{8\pi r} \ddot{D}_{\alpha\beta} \;, \quad
    \bar{f}_{(0\alpha)} = -
    \frac{x^{\beta}}{r} 
      \frac{\kappa}{8\pi r} \ddot{D}_{\alpha\beta} \;, \\
    \bar{f}_{(00)} &=
    \frac{\kappa}{4\pi r} \sum_{a=1}^{N} m_{a}c^{2} +
      \frac{x^{\alpha}x^{\beta}}{r^{2}}
      \frac{\kappa}{8\pi r} \ddot{D}_{\alpha\beta} \;.
  \end{align}
\end{subequations}
Here,
\begin{math}
  D_{\alpha\beta}
\end{math}
is the quadrupole moment of the mass distribution defined by
\begin{equation}
  \label{eq:quadrupole_moment}
  D_{\alpha\beta} \stackrel{\mathrm{def}}{=}
  \sum_{a = 1}^{N} m_{a} \xi^{\alpha}_{a} \xi^{\beta}_{a} \;,
\end{equation}
where we have defined
\begin{math}
  \ddot{D}_{\alpha\beta} \stackrel{\mathrm{def}}{=}
  \partial^{2} D_{\alpha\beta} / \partial^{2} t.
\end{math}
The wave form given by Eqs.~(\ref{eq:quadrupole_formula}) and
(\ref{eq:quadrupole_moment}) is the same as the corresponding wave
form in GR, which is consistent with the result given in
Ref.~\citen{schweizer.straumann.wipf:1980}.

\section{Emission rates of energy-momentum and angular momentum}
\label{sec:emission_rates}

In this section, we examine time averages of emission rates of the
energy-momentum and angular momentum carried by the quadrupole
radiation given by Eq.~(\ref{eq:quadrupole_formula}).

In ENGR, for the asymptotic conditions (\ref{eq:asymptotic_vierbein})
and (\ref{eq:asymptotic_vierbein.ii}), we have four quantities that
are conserved as long as they are finite, the dynamical
energy-momentum
\begin{math}
  M_{k},
\end{math}
\lq\lq{}spin'' angular momentum
\begin{math}
  S_{kl},
\end{math}
canonical energy-momentum
\begin{math}
  M^{c}_{\hspace{0.7ex}\mu},
\end{math}
and \lq\lq{}extended orbital angular momentum''
\begin{math}
  L_{\mu}^{\hspace{0.7ex}\nu}
\end{math}~\cite{kawai.toma:1991}. \emph{The dynamical energy momentum
  does not depend on the asymptotic form of the Higgs-type field
  $\psi^{k}$ explicitly,} but the quantities
\begin{math}
  S_{kl}, M^{c}_{\hspace{0.7ex}\mu}
\end{math}
and
\begin{math}
  L_{\mu}^{\hspace{0.7ex}\nu}
\end{math}
depend on the asymptotic form.  With this in mind, following
Ref.~\citen{kawai.toma:1991}, we examine the case in which the
asymptotic form of the Higgs-type field is
\begin{displaymath}
  \psi^{k} \simeq e^{(0)k}_{\hspace{2.8ex}\mu} x^{\mu} +
  \psi^{(0)k} \;,
\end{displaymath}
with
\begin{math}
  \psi^{(0)k}
\end{math}
constant. We also consider the slightly generalized case in which
\begin{displaymath}
  \psi^{k} \simeq \rho e^{(0)k}_{\hspace{2.8ex}\mu} x^{\mu} +
  \psi^{(0)k}.
\end{displaymath}
The asymptotic form of the function $\rho$ is determined in
\S\ref{sec:asymptotic_form_case.ii}.

\subsection{The case $\psi^{k} \simeq 
  e^{(0)k}_{\hspace{2.8ex}\mu}x^{\mu} + \psi^{(0)k}$}
\label{sec:asymptotic_form_case.i}

\subsubsection{Dynamical energy-momentum loss}
\label{sec:energy_momentum}

In order to evaluate the emission rate of the dynamical
energy-momentum, we integrate the differential
conservation law (\ref{eq:differential_conservation_law_emd}) over
a large solid sphere $V$ with radius $r$, yielding
\begin{equation}
  \label{eq:integration_continuous_equation}
  \partial_{0}\int_{V}
    {}^{\mathrm{tot}}\bm{T}_{k}^{\hspace{0.7ex}0}d^{3}x =
  - \int_{S} {}^{\mathrm{tot}}\bm{T}_{k}^{\hspace{0.7ex}\alpha}
  r^{2} n^{\alpha} d\Omega \;,
\end{equation}
where $S$ and $d\Omega$ represent the two-dimensional surface of $V$
and the differential solid angle, respectively. Also, $n^{\alpha}$
stands for the unit radial vector defined by
\begin{math}
  n^{\alpha} \stackrel{\mathrm{def}}{=} x^{\alpha} / r.
\end{math}
Taking into account the fact that the energy-momentum density of point
masses vanishes for very large $r$, we can rewrite the r.h.s.\ of
Eq.~(\ref{eq:integration_continuous_equation}) as
\begin{equation}
  \label{eq:dynamical_energy_momentum_loss}
  - \int_{S} {}^{\mathrm{tot}}\bm{T}_{k}^{\hspace{0.7ex}\alpha}
  r^{2} n^{\alpha} d\Omega =
  - \int_{S} {}^{G}\bm{T}_{k}^{\hspace{0.7ex}\alpha}
  r^{2} n^{\alpha} d\Omega \;.
\end{equation}
The density 
\begin{math}
  {}^{G}\bm{T}_{k}^{\hspace{0.7ex}\mu}
\end{math}
takes, up to terms of order
\begin{math}
  O(1 / r^{2}),
\end{math}
the form
\begin{equation}
  \label{eq:dynamical_emdensity_leading}
  {}^{G}\bm{T}_{k}^{\hspace{0.7ex}\mu} =
  e^{(0)\nu}_{\hspace{2.8ex}k}
    {}^{G}T_{\hspace{2.1ex}\nu}^{(0)\hspace{0.7ex}\mu} \;,
\end{equation}
with
\begin{equation}
  \label{eq:dynamical_emdensity_leading_explicit}
  {}^{G}T^{(0)\hspace{0.7ex}\mu}_{\hspace{2.1ex}\nu}
  \stackrel{\mathrm{def}}{=}
  \frac{1}{\kappa} \left(
  \partial^{\mu}\bar{f}^{(\rho\sigma)}
    \partial_{\nu}\bar{f}_{(\rho\sigma)} -
  \frac{1}{2}\partial^{\mu}\bar{f} \partial_{\nu}\bar{f} \right).
\end{equation}
Using the solution (\ref{eq:quadrupole_formula}), and averaging over
one period of motion of the system of point masses, we obtain
\begin{equation}
  \label{eq:energy_loss}
  - \left\langle \frac{dE}{dt} \right\rangle =
  \frac{G}{5c^{5}} \left\langle
    \dddot{D}_{\alpha\beta} \dddot{D}_{\alpha\beta} -
    \frac{1}{3}\dddot{D}_{\alpha\alpha}
    \dddot{D}_{\beta\beta} \right\rangle,
\end{equation}
for a total energy 
\begin{math}
  E \stackrel{\mathrm{def}}{=} - M_{(0)}
\end{math}
of the system, where
\begin{math}
  \langle \cdots \rangle
\end{math}
denotes the operation of averaging over one period of motion of
the system of the point masses, and we have set
\begin{math}
  e^{(0)k}_{\hspace{2.8ex}\mu} = \delta^{k}_{\hspace{0.7ex}\mu}
\end{math}
for simplicity.  Also, we obtain
\begin{equation}
  \label{eq:momentum_loss}
  \frac{dM_{a}}{dt} = 0.
\end{equation}
It is worth mentioning that the quantity
\begin{math}
  {}^{G}T^{(0)\hspace{0.7ex}\nu}_{\hspace{2.1ex}\mu}
\end{math}
is identical to the r.h.s.\ of
Eq.~(\ref{eq:weak_emd_gravity_leading.GR}), and the time average of
the emission rate of the dynamical energy-momentum, which is given by
Eqs.~(\ref{eq:energy_loss}) and (\ref{eq:momentum_loss}), is identical
to that of the canonical energy-momentum in GR.

\subsubsection{\lq\lq{}Spin'' angular momentum loss}
\label{sec:spin}

The density
\begin{math}
  {}^{G}\bm{S}_{kl}^{\hspace{1.4ex}\mu}
\end{math}
has the expression 
\begin{align}
  \label{eq:weak_samd_gravity_ENGR}
  {}^{G}\bm{S}_{kl}^{\hspace{1.4ex}\mu} &=
  2 \eta_{km} \eta_{ln} e^{(0)m}_{\hspace{2.8ex}[\rho}
  e^{(0)n}_{\hspace{2.8ex}\sigma]} \psi^{\rho}
    (- g)t^{\sigma\mu}_{\mathrm{LL}} \notag \\
  & - 2 e^{(0)\nu}_{\hspace{2.8ex}[k} \eta_{l]m} 
  e^{(0)m}_{\hspace{2.8ex}\tau} \left[
    Z^{(1)\hspace{0.7ex}\mu}_{\hspace{2.1ex}\nu} \psi^{\tau} +
    \eta_{\nu\sigma} (Z^{(2)\mu\sigma\tau} -
    Z^{(3)\mu\lambda\sigma}\psi^{\tau}_{\hspace{0.7ex},\lambda})
  \right], \notag \\
\end{align}
where we have defined
\begin{align}
  \psi^{\rho} &\stackrel{\mathrm{def}}{=}
  e^{(0)\rho}_{\hspace{2.8ex}k}\psi^{k} \;, \\
  \label{eq:definition_Z.i}
  \kappa Z^{(1)\hspace{0.7ex}\mu}_{\hspace{2.1ex}\nu} 
  &\stackrel{\mathrm{def}}{=}
  - \frac{1}{2}\delta_{\nu}^{\hspace{0.7ex}\mu}
    \partial^{\sigma}\bar{f}^{(\lambda\rho)}
    \partial_{\rho}\bar{f}_{(\lambda\sigma)}
  - \partial^{\sigma}\bar{f}^{(\rho\mu)}
    \partial_{\sigma}\bar{f}_{(\rho\nu)}
  + \partial^{\sigma}\bar{f}^{(\rho\mu)}
    \partial_{\nu}\bar{f}_{(\rho\sigma)} \notag \\
  & + \partial^{\mu}\bar{f}^{(\rho\sigma)}
    \partial_{\sigma}\bar{f}_{(\rho\nu)}
  - \frac{1}{2} \partial^{\sigma}\bar{f}^{(\rho\mu)}
    \eta_{\rho\nu}\partial_{\sigma}\bar{f}
  + \frac{1}{2} \partial^{\mu}\bar{f}_{(\nu\sigma)}
    \partial^{\sigma}\bar{f} \;, \\
  \label{eq:definition_Z.ii}
  \kappa Z^{(2)\mu\sigma\tau} &\stackrel{\mathrm{def}}{=}
  \left(
    \partial^{\lambda}\bar{f}^{(\sigma\mu)} -
    \partial^{\mu}\bar{f}^{(\sigma\lambda)} \right) \left(
    \delta^{\tau}_{\hspace{0.7ex}\lambda} + 
    \eta^{\tau\rho}\bar{f}_{(\rho\lambda)} -
    \frac{1}{2}\delta^{\tau}_{\hspace{0.7ex}\lambda}\bar{f} 
  \right), \\
  \label{eq:definition_Z.iii}
  \kappa Z^{(3)\mu\lambda\sigma} &\stackrel{\mathrm{def}}{=}
  \partial^{\lambda}\bar{f}^{(\sigma\mu)} -
  \partial^{\mu}\bar{f}^{(\sigma\lambda)} \;.
\end{align}
Also,
\begin{math}
  t^{\sigma\mu}_{\mathrm{LL}}
\end{math}
denotes the energy-momentum density introduced by Landau and Lifshitz,
whose explicit form is given in Appendix~\ref{sec:einstein_theory}.
Using a method similar to that employed in
\S\ref{sec:energy_momentum}, we obtain
\begin{equation}
  \label{eq:spin_angular_momentum_loss}
  \partial_{0} \int_{V} 
  {}^{\mathrm{tot}}\bm{S}_{kl}^{\hspace{1.4ex}0} d^{3}x = -
  \int_{S} {}^{G}\bm{S}_{kl}^{\hspace{1.4ex}\alpha}
  r^{2}n^{\alpha} d\Omega \;.
\end{equation}
In order to estimate the r.h.s.\ of
Eq.~(\ref{eq:spin_angular_momentum_loss}), we set
\begin{equation}
  \label{eq:asymptotic_higgs_type_field}
  \psi^{\mu} =
  x^{\mu} + \psi^{(0)\mu} + \tilde{\psi}^{\mu} \;,
\end{equation}
where
\begin{math}
  \psi^{(0)\mu}
\end{math}
is a constant.  From Eqs.~(\ref{eq:quadrupole_formula}) and
(\ref{eq:weak_samd_gravity_ENGR})--(%
\ref{eq:asymptotic_higgs_type_field}),
we can show that\footnote{Note that we have distinguished between
  Latin and Greek indices here. (See
  footnote~\ref{footnote:distinction_between_indices} on page
  \pageref{footnote:distinction_between_indices}.)}
\begin{align}
  \label{eq:spin_am_time_space_loss}
  -\left\langle \frac{dS_{(0)a}}{dt} \right\rangle &=
  - \psi^{(0)a} \frac{G}{5c^{6}} \left\langle
    \dddot{D}_{\alpha\beta} \dddot{D}_{\alpha\beta} -
    \frac{1}{3} \dddot{D}_{\alpha\alpha} \dddot{D}_{\beta\beta}
  \right\rangle \notag \\
  &= - 2\psi^{(0)}_{\hspace{2.1ex}[a} \left\langle 
    \frac{dM_{(0)]}}{dt} \right\rangle
\end{align}
if
\begin{equation}
  \label{eq:asymptotic_higgs_condition.i}
  \lim_{r \to \infty} \tilde{\psi}^{\mu} = 0, \quad
  \lim_{r \to \infty}
    \tilde{\psi}^{\mu}_{\hspace{0.7ex},0} = 0, \quad
  \lim_{r \to \infty} 
    r\tilde{\psi}^{\mu}_{\hspace{0.7ex},\beta} = 0,
\end{equation}
where again we have set
\begin{math}
  e^{(0)k}_{\hspace{2.8ex}\mu} = \delta^{k}_{\hspace{0.7ex}\mu}.
\end{math}
Also, we have
\begin{equation}
  \label{eq:spin_am_space_space_loss}
  - \left\langle \frac{dS_{ab}}{dt} \right\rangle =
  \frac{2G}{5c^{5}} 
  \left\langle
    \ddot{D}_{a\gamma} \dddot{D}_{b\gamma} -
    \ddot{D}_{b\gamma} \dddot{D}_{a\gamma} \right\rangle
\end{equation}
if
\begin{align}
  \label{eq:asymptotic_higgs_condition.ii}
  \lim_{r \to \infty} \tilde{\psi}^{\alpha} &= 0, \quad
  \lim_{r \to \infty}
    \tilde{\psi}^{\alpha}_{\hspace{0.7ex},0} = 0, \quad
  \lim_{r \to \infty} 
    r\tilde{\psi}^{\alpha}_{\hspace{0.7ex},\beta} = 0.
\end{align}
The time average of the emission rate of the \lq\lq{}spin'' angular
momentum
\begin{math}
  S_{ab}
\end{math}
is the same as that of the space-space component of the angular
momentum in GR.

\subsubsection{Canonical energy-momentum and 
  orbital angular momentum \\ losses}
\label{sec:generators.coordinate}

In the weak-field approximation, the canonical energy-momentum density
of the gravitational field
\begin{math}
  {}^{G}\widetilde{\bm{T}}_{\mu}^{\hspace{0.7ex}\nu}
\end{math}
becomes
\begin{equation}
  \label{eq:weak_canonical_emd.ENGR}
  {}^{G}\widetilde{\bm{T}}_{\mu}^{\hspace{0.7ex}\nu} =
  \eta_{\rho\sigma} Z^{(3)\nu\lambda\sigma}
    \psi^{\rho}_{\hspace{0.7ex},\lambda\mu} -
  {}^{G}T_{\hspace{2.1ex}\lambda}^{(0)\hspace{0.7ex}\nu}
    \psi^{\lambda}_{\hspace{0.7ex},\mu} \;.
\end{equation}
Then, using Eqs.~(\ref{eq:quadrupole_formula}),
(\ref{eq:definition_Z.iii}) and
(\ref{eq:asymptotic_higgs_type_field}), we can show that
\begin{equation}
  \label{eq:canonical_energy_momentum_loss}
  \frac{dM^{c}_{\hspace{0.7ex}\mu}}{dt} = 0
\end{equation}
if the conditions
\begin{equation}
  \label{eq:asymptotic_higgs_condition.iii}
  \lim_{r \to \infty}
    \tilde{\psi}^{\mu}_{\hspace{0.7ex},\nu} = 0, \quad
  \lim_{r \to \infty}
    \tilde{\psi}^{\mu}_{\hspace{0.7ex},00} = 0, \quad
  \lim_{r \to \infty}
    r\tilde{\psi}^{\mu}_{\hspace{0.7ex},\beta\nu} = 0
\end{equation}
are satisfied.

Finally, we examine the \lq\lq{}extended orbital angular momentum.''
In the weak-field approximation, the \lq\lq{}extended orbital angular
momentum'' density of the gravitational field
\begin{math}
  {}^{G}\widetilde{\bm{M}}_{\mu}^{\hspace{0.7ex}\nu\lambda}
\end{math}
becomes
\begin{equation}
  \label{eq:weak_extended_oamd_gravity.ENGR}
  {}^{G}\widetilde{\bm{M}}_{\mu}^{\hspace{0.7ex}\nu\lambda} =
  2Z^{(4)\hspace{0.7ex}\nu\lambda}_{\hspace{2.1ex}\mu} -
  2Z^{(3)\lambda\nu\sigma} \eta_{\sigma\rho}
    \psi^{\rho}_{\hspace{0.7ex},\mu} -
  2x^{\nu} Z^{(3)\lambda\tau\sigma} \eta_{\rho\sigma}
    \psi^{\rho}_{\hspace{0.7ex},\tau\mu} +
  2x^{\nu} Z^{(5)\hspace{0.7ex}\lambda}_{\hspace{2.1ex}\sigma}
    \psi^{\sigma}_{\hspace{0.7ex},\mu} \;,
\end{equation}
where we have defined
\begin{align}
  \label{eq:definition_Z.iv}
  \kappa Z^{(4)\hspace{0.7ex}\nu\lambda}_{\hspace{2.1ex}\mu}
  &\stackrel{\mathrm{def}}{=}
  \left(\partial^{\nu}\bar{f}^{(\sigma\lambda)} -
    \partial^{\lambda}\bar{f}^{(\sigma\nu)} \right) \left(
      \eta_{\sigma\mu} + \bar{f}_{(\sigma\mu)} -
      \frac{1}{2}\eta_{\sigma\mu}\bar{f} \right) \notag \\
  & + x^{\nu} \biggl[
    \delta_{\mu}^{\hspace{0.7ex}\lambda} \biggl(
      \frac{1}{2}\partial^{\sigma}\bar{f}^{(\tau\rho)}
      \partial_{\sigma}\bar{f}_{(\tau\rho)}
      - \frac{1}{2}\partial^{\sigma}\bar{f}^{(\tau\rho)}
      \partial_{\rho}\bar{f}_{(\tau\sigma)}
      - \frac{1}{4} \partial^{\sigma}\bar{f}
      \partial_{\sigma}\bar{f} \biggr) \notag \\
  & \quad + \left(\partial^{\rho}\bar{f}^{(\sigma\lambda)} -
    \partial^{\lambda}\bar{f}^{(\sigma\rho)} \right) \left(
    \partial_{\mu}\bar{f}_{(\sigma\rho)} -
    \frac{1}{2}\eta_{\sigma\rho}\partial_{\mu}\bar{f} \right)
  \biggr], \\
  \label{eq:definition_Z.v}
  \kappa Z^{(5)\hspace{0.7ex}\lambda}_{\hspace{2.1ex}\sigma}
  &\stackrel{\mathrm{def}}{=}
  \delta_{\sigma}^{\hspace{0.7ex}\lambda} \left(
    - \frac{1}{2}\partial^{\xi}\bar{f}^{(\tau\rho)}
      \partial_{\xi}\bar{f}_{(\tau\rho)}
    + \frac{1}{2}\partial^{\xi}\bar{f}^{(\tau\rho)}
      \partial_{\rho}\bar{f}_{(\tau\xi)}
    + \frac{1}{4} \partial^{\rho}\bar{f}
      \partial_{\rho}\bar{f} \right) \notag \\
  &
    + \partial^{\lambda}\bar{f}^{(\rho\tau)}
      \partial_{\sigma}\bar{f}_{(\rho\tau)}
    - \frac{1}{2} \partial^{\lambda}\bar{f}\partial_{\sigma}\bar{f} 
    + \partial^{\tau}\bar{f}^{(\rho\lambda)}
      \partial_{\tau}\bar{f}_{(\rho\sigma)}
    - \partial^{\tau}\bar{f}^{(\rho\lambda)}
      \partial_{\sigma}\bar{f}_{(\rho\tau)} \notag \\
  &
    - \partial^{\lambda}\bar{f}^{(\rho\tau)}
      \partial_{\tau}\bar{f}_{(\rho\sigma)}
    - \frac{1}{2}\partial^{\tau}\bar{f}^{(\rho\lambda)}
      \eta_{\rho\sigma}\partial_{\tau}\bar{f}
    + \frac{1}{2}\partial^{\lambda}\bar{f}_{(\sigma\tau)}
      \partial^{\tau}\bar{f} \;.
\end{align}
As can be shown by using Eqs.~(\ref{eq:quadrupole_formula}),
(\ref{eq:asymptotic_higgs_type_field}),
(\ref{eq:weak_extended_oamd_gravity.ENGR}), (\ref{eq:definition_Z.iv})
and (\ref{eq:definition_Z.v}), we have the relations
\begin{align}
  \label{eq:extended_angular_momentum_loss}
  \frac{dL_{\mu}^{\hspace{0.7ex}0}}{dt} &= 0, \quad
  \frac{dL_{0}^{\hspace{0.7ex}\alpha}}{dt} = 0, \\
  \left\langle 
    \frac{dL_{\alpha}^{\hspace{0.7ex}\beta}}{dt} \right\rangle &=
  - \frac{G}{3c^{5}} \left\langle
    \ddot{D}_{\alpha\gamma}\dddot{D}_{\beta\gamma} +
    \ddot{D}_{\alpha\gamma}\dddot{D}_{\beta\gamma} -
    \ddot{D}_{\gamma\gamma}\dddot{D}_{\alpha\beta} \right\rangle
\end{align}
if the conditions
\begin{equation}
  \label{eq:asymptotic_higgs_condition.iv}
  \lim_{r \to \infty}
    r\tilde{\psi}^{\mu}_{\hspace{0.7ex},\nu} = 0, \quad
  \lim_{r \to \infty}
    r\tilde{\psi}^{\mu}_{\hspace{0.7ex},00} = 0, \quad
  \lim_{r \to \infty}
    r^{2}\tilde{\psi}^{\mu}_{\hspace{0.7ex},\beta\nu} = 0
\end{equation}
are satisfied. The emission rate of the component
\begin{math}
  L_{\alpha}^{\hspace{0.7ex}\beta}
\end{math}
is finite. However, for the antisymmetric part of
\begin{math}
  L_{\alpha\beta},
\end{math}
which is the three-dimensional orbital angular momentum, we have
\begin{equation}
  \label{eq:antisymmetric_part_ss_orbital_angulr_momentum}
  \frac{dL_{[\alpha\beta]}}{dt} = 0.
\end{equation}

To summarize, the emission rates
\begin{math}
  dM^{c}_{\hspace{0.7ex}\mu} / dt
\end{math}
and
\begin{math}
  dL_{[\mu\nu]} / dt
\end{math}
are both vanishing if the condition
(\ref{eq:asymptotic_higgs_condition.iv}) is satisfied.\footnote{Note
  that $L_{[\mu\nu]}$ corresponds to the generator of the Lorentz
  coordinate transformations and that it possibly represents the
  four-dimensional orbital angular momentum.}

\subsection{The case $\psi^{k} \simeq 
  \rho e^{(0)k}_{\hspace{2.8ex}\mu}x^{\mu} + \psi^{(0)k}$}
\label{sec:asymptotic_form_case.ii}

Also in this case, the expressions (\ref{eq:energy_loss}) and 
(\ref{eq:momentum_loss}) for the time average of the emission rate of
the dynamical energy-momentum hold, as seen from
Eq.~(\ref{eq:dynamical_emdensity_leading}).

We express $\psi^{\mu }$ as 
\begin{equation}
  \label{eq:asymptotic_higgs_type_field.ii}
  \psi^{\mu} = 
  \rho(r,t)x^{\mu} + \psi^{(0)\mu} + \tilde{\psi}^{\mu} \;.
\end{equation}
Then, the expression (\ref{eq:spin_am_time_space_loss}) for the time
average of the emission rate of the time-space component of the
\lq\lq{}spin'' angular momentum holds if the condition
(\ref{eq:asymptotic_higgs_condition.i}) is satisfied by
$\tilde{\psi}^{\mu}$ in Eq.~(\ref{eq:asymptotic_higgs_type_field.ii}).
For the space-space component
\begin{math}
  S_{ab},
\end{math}
using Eqs.~(\ref{eq:quadrupole_formula}) and
(\ref{eq:weak_samd_gravity_ENGR})--(%
\ref{eq:spin_angular_momentum_loss}), we find
\begin{equation}
  \label{eq:spin_am_space_space_loss.ii}
  - \left\langle \frac{dS_{ab}}{dt} \right\rangle =
  \frac{2G}{5c^{5}} \frac{3 + \rho_{c}}{4}
  \left\langle
    \ddot{D}_{a\gamma} \dddot{D}_{b\gamma} -
    \ddot{D}_{b\gamma} \dddot{D}_{a\gamma} \right\rangle
\end{equation}
if the condition (\ref{eq:asymptotic_higgs_condition.ii}) and
\begin{equation}
  \label{eq:asymptotic_higgs_condition.v}
  \lim_{r \to \infty} \rho = \rho_{c} \;,
\end{equation}
with $\rho_{c}$ constant, are satisfied by $\rho$ and
$\tilde{\psi}^{\mu}$ in Eq.~(\ref{eq:asymptotic_higgs_type_field.ii}).

From Eqs.~(\ref{eq:quadrupole_formula}),
(\ref{eq:dynamical_emdensity_leading_explicit}),
(\ref{eq:definition_Z.iii}), (\ref{eq:weak_canonical_emd.ENGR}) and
(\ref{eq:asymptotic_higgs_type_field.ii}), we have the relations
\begin{align}
  \label{eq:canonical_energy_loss}
  \left\langle \frac{dM^{c}_{\hspace{0.7ex}0}}{dt} \right\rangle &=
  \frac{G}{5c^{5}} (1 - \rho_{c})\left\langle
    \dddot{D}_{\alpha\beta} \dddot{D}_{\alpha\beta} -
    \frac{1}{3}\dddot{D}_{\alpha\alpha}
    \dddot{D}_{\beta\beta} \right\rangle, \\
  \label{eq:canonical_momentum_loss}
  \frac{dM^{c}_{\hspace{0.7ex}\alpha}}{dt} &= 0
\end{align}
if the condition (\ref{eq:asymptotic_higgs_condition.iii}) and
\begin{equation}
  \label{eq:asymptotic_higgs_condition.vi}
  \lim_{r \to \infty} \rho = \rho_{c} \;, \quad 
  \lim_{r \to \infty} r\rho_{,0} = 0, \quad
  \lim_{r \to \infty} r\rho_{,00} = 0
\end{equation}
are satisfied.

Under the condition (\ref{eq:asymptotic_higgs_condition.v}), the time
average of the emission rate of the \lq\lq{}extended orbital angular
momentum'' diverges. However, using Eqs.~(\ref{eq:quadrupole_formula}),
(\ref{eq:weak_extended_oamd_gravity.ENGR}),
(\ref{eq:definition_Z.iv}), (\ref{eq:definition_Z.v}) and
(\ref{eq:asymptotic_higgs_type_field.ii}), we can show that
\begin{align}
  \frac{dL_{[0\alpha]}}{dt} &= 0, \\
  \label{eq:orbital_angular_momentum_loss}
  - \left\langle \frac{dL_{[\alpha\beta]}}{dt} \right\rangle &=
  \frac{2G}{5c^{5}}\frac{1 - \rho_{c}}{4}
  \left\langle
    \ddot{D}_{a\gamma} \dddot{D}_{b\gamma} -
    \ddot{D}_{b\gamma} \dddot{D}_{a\gamma} \right\rangle
\end{align}
if the conditions (\ref{eq:asymptotic_higgs_condition.iv}) and
(\ref{eq:asymptotic_higgs_condition.v}) are satisfied.

For a space-time satisfying the asymptotic conditions
(\ref{eq:asymptotic_vierbein}) and (\ref{eq:asymptotic_vierbein.ii}),
the sum
\begin{math}
  S_{kl} + e^{(0)\mu}_{\hspace{2.8ex}k} 
    e^{(0)\nu}_{\hspace{2.8ex}l} L_{[\mu\nu]}
\end{math}
is well-defined and conserved for the case
\begin{math}
  \psi^{k} \simeq \rho_{c}e^{(0)k}_{\hspace{2.8ex}\mu}x^{\mu} +
  \psi^{(0)k} \; (\rho_{c} \neq 1), 
\end{math}
as described in Ref.~\citen{kawai.toma:1991}. 
In the case under consideration, we have
\begin{align}
  \label{eq:spin.orbital_am_time_space_loss}
  - \left\langle \frac{d}{dt}(S_{(0)a} + L_{[0a]}) \right\rangle &=
  - 2\psi^{(0)}_{\hspace{2.1ex}[a} \left\langle 
    \frac{dM_{(0)]}}{dt} \right\rangle, \\
  \label{eq:spin.orbital_am_space_space_loss}
  - \left\langle \frac{d}{dt}\left(
    S_{ab} + L_{[ab]} \right) \right\rangle &=
  \frac{2G}{5c^{5}} \left\langle
    \ddot{D}_{a\gamma} \dddot{D}_{b\gamma} -
    \ddot{D}_{b\gamma} \dddot{D}_{a\gamma} \right\rangle
\end{align}
if the conditions
\begin{equation}
  \label{eq:asymptotic_higgs_condition.vii}
  \lim_{r \to \infty} \tilde{\psi}^{\mu} = 0, \quad
  \lim_{r \to \infty} 
    r \tilde{\psi}^{\mu}_{\hspace{0.7ex},\nu} = 0, \quad
  \lim_{r \to \infty}
    r \tilde{\psi}^{\mu}_{\hspace{0.7ex},00} = 0, \quad
  \lim_{r \to \infty}
    r^{2} \tilde{\psi}^{\mu}_{\hspace{0.7ex},\beta\nu} = 0
\end{equation}
are satisfied. Here, again, we have set
\begin{math}
  e^{(0)\mu}_{\hspace{2.8ex}k} = \delta^{\mu}_{\hspace{0.7ex}k}.
\end{math}

\section{Summary and discussion}
\label{sec:summary}

In an extended, new form of general relativity, we have examined
energy-momentum and angular momentum carried by gravitational wave
radiated from a system of Newtonian point masses in a weak-field
approximation.  The results are summarized as follows.
\begin{enumerate}
\item\label{summary:wave_form} The form of the gravitational wave is
  identical to the corresponding wave form in GR, which is consistent
  with the result in Ref.~\citen{schweizer.straumann.wipf:1980}.
\item\label{summary:energy_momentum_plane_wave} The average value of
  $\langle {}^{G}\bm{T}_{l}^{\hspace{0.7ex}\mu} \rangle$ for the
  dynamical energy-momentum of a plane wave is obtained from
  Eq.~(\ref{eq:wave_length_average}), and it is the same as that of
  the corresponding canonical energy-momentum in GR~\cite{weinberg}.
\item\label{summary:dynamical_energy_momentum} The dynamical
  energy-momentum density 
  \begin{math}
    {}^{G}\bm{T}_{k}^{\hspace{0.7ex}\mu}
  \end{math}
  takes, up to order
  \begin{math}
    O(1 / r^{2}),
  \end{math}
  the form given by Eq.~(\ref{eq:dynamical_emdensity_leading}) with
  Eq.~(\ref{eq:dynamical_emdensity_leading_explicit}), and this is
  essentially the same as the corresponding expression
  (\ref{eq:weak_emd_gravity_leading.GR}) for the canonical
  energy-momentum density in GR, and the time average of the
  energy-momentum emission rate is given by
  Eqs.~(\ref{eq:energy_loss}) and (\ref{eq:momentum_loss}), which
  is identical to that in GR.
\item\label{summary:spin_angular_momentum} The time average of the
  emission rate of the \lq\lq{}spin'' angular momentum is given by
  Eqs.~(\ref{eq:spin_am_time_space_loss}) and
  (\ref{eq:spin_am_space_space_loss}) if the conditions
  (\ref{eq:asymptotic_higgs_condition.i}) and
  (\ref{eq:asymptotic_higgs_condition.ii}) for the form
  (\ref{eq:asymptotic_higgs_type_field}) of the Higgs-type field are
  satisfied. The expression (\ref{eq:spin_am_space_space_loss}) is the
  same as the corresponding expression for the angular momentum in GR.
\item\label{summary:canonical_energy_orbital_angular_momentum} The
  emission rates of both the canonical energy-momentum and the orbital
  angular momentum vanish if the conditions
  (\ref{eq:asymptotic_higgs_condition.iii}) and
  (\ref{eq:asymptotic_higgs_condition.iv}) for $\psi^{k}$ given by the
  expression (\ref{eq:asymptotic_higgs_type_field}) are both
  satisfied.
\item\label{summary:another_case_asymptotic_higgs_type_field} Under
  the condition (\ref{eq:asymptotic_higgs_condition.v}) for the form
  (\ref{eq:asymptotic_higgs_type_field.ii}) of the Higgs-type field,
  the time average of the emission rates of the \lq\lq{}spin'' angular
  momentum and the canonical energy-momentum depend on the constant
  $\rho_{c}$. Moreover, the time average of the emission rate of the
  \lq\lq{}extended orbital angular momentum'' diverges. However, the
  time average of the emission rate of the sum~\cite{kawai.toma:1991}
  \begin{math}
    S_{kl} + e^{(0)\mu}_{\hspace{2.8ex}k}
      e^{(0)\nu}_{\hspace{2.8ex}l} L_{[\mu\nu]}
  \end{math}
  is finite, and its space-space component is identical to the
  corresponding expression for the angular momentum in GR if the
  condition (\ref{eq:asymptotic_higgs_condition.vii}) is satisfied.
\end{enumerate}

Finally, we would like to add the following:
\begin{enumerate}
\renewcommand{\theenumi}{\Alph{enumi}}
\item\label{discussion:comparing_with_asymptotically_flat_space} As we
  have stated repeatedly, the dynamical energy-momentum and
  \lq\lq{}spin'' angular momentum are generators of \emph{internal}
  translations and \emph{internal} $SL(2,C)$ transformations. The
  former does not depend on the non-dynamical field $\psi^{k}$
  explicitly, but the latter does. For vierbeins possessing
  asymptotic forms satisfying Eqs.~(\ref{eq:asymptotic_vierbein}) and
  (\ref{eq:asymptotic_vierbein.ii}), they give the total energy-momentum
  and \emph{total} (=\emph{spin}+\emph{orbital}) angular momentum of
  the system when the field $\psi^{k}$ is chosen as
  \begin{math}
    \psi^{k} = 
    e^{(0)k}_{\hspace{2.8ex}\mu}x^{\mu} + \psi^{(0)k} +
    O(1/r^{\beta }).
  \end{math}
  The generator of the affine \emph{coordinate} transformations,
   contrastingly, vanishes. The results summarized
  in~\ref{summary:energy_momentum_plane_wave}--\ref{%
    summary:canonical_energy_orbital_angular_momentum} above are
  consistent with this. The discussion in
  Appendix~\ref{sec:problem_three} gives further support to the choice
  of the asymptotic form of $\psi^{k}$ given by
  Eq.~(\ref{eq:asymptotic_higgs_type_field}) satisfying the condition
  (\ref{eq:asymptotic_higgs_condition.ii}).
  
\item\label{discussion:higgs_type_field} The \lq\lq{}spin'' angular
  momentum depends on the Higgs-type field $\psi^{k}$, and it is
  meaningful if $\psi^{k}$ satisfies a suitable condition requiring
  this field to be the same as the Minkowskian coordinates on the
  boundary of a sphere of infinite radius, as is known from
  \S\ref{sec:emission_rates} and from
  Refs.~\citen{kawai.toma:1991,kawai:1988,%
    kawai.saitoh:1989a,kawai.saitoh:1989b}.  The field $\psi^{k}$
  behaves like a Minkowskian coordinate system under \emph{internal}
  $\overline{\mbox{Poincar\'{e}}}$ transformations, and its existence
  is a necessary consequence of the structure of the group
  $\bar{P}_{0}$ and a basic postulate regarding the space-time. Also,
  this field is closely related to the existence of the spinor
  structure~\cite{kawai:1986}.  However, the physical and geometrical
  meaning of this field has not yet been fully clarified.
  
\item\label{discussion:choice_of_independent_field_variables} In
  considering the energy-momentum and angular momentum, there are two
  possibilities in choosing the set of independent field
  variables~\cite{kawai.toma:1991,kawai:2000}, i.e., the set
  \begin{math}
    \{\psi^{k}, A^{k}_{\hspace{0.7ex}\mu}, \phi^{A}\}
  \end{math}
  and the set
  \begin{math}
    \{\psi^{k}, e^{k}_{\hspace{0.7ex}\mu}, \phi^{A}\}.
  \end{math}
  In the present paper, we have employed 
  \begin{math}
    \{\psi^{k}, A^{k}_{\hspace{0.7ex}\mu}, \phi^{A}\}
  \end{math}
  as the set of independent field variables, because this choice is
  superior to the other, as shown Refs.~\citen{kawai.toma:1991} and
  \citen{kawai:2000}.\footnote{It should be noted that when
    $\{\psi^{k}, A^{k}_{\hspace{0.7ex}\mu}, \phi^{A}\}$ is employed as
    the set of independent field variables, the Lagrangian $L$ is
    considered to have an explicit $\psi^{k}$ dependence, because $L$
    is a function of $e^{k}_{\hspace{0.7ex}\mu} =
    \psi^{k}_{\hspace{0.7ex},\mu} + A^{k}_{\hspace{0.7ex}\mu}, \;
    \phi^{A}$ and their derivatives.}

\item\label{discussion:asymptotic_form_higgs_type_field}
  \begin{enumerate}
  \item The asymptotic conditions
    (\ref{eq:asymptotic_higgs_condition.iv}) and
    (\ref{eq:asymptotic_higgs_condition.v}) for the field $\psi^{k}$
    are stronger than the corresponding conditions for vierbeins whose
    asymptotic forms satisfy Eqs.~(\ref{eq:asymptotic_vierbein}) and
    (\ref{eq:asymptotic_vierbein.ii})~\cite{kawai.toma:1991}.  This is
    natural, because vierbeins in which gravitational waves propagate
    do not satisfy the conditions (\ref{eq:asymptotic_vierbein}) and
    (\ref{eq:asymptotic_vierbein.ii}) in general.

  \item For the form (\ref{eq:asymptotic_higgs_type_field.ii}) of a
    Higgs-type field, the time average of the emission rate of the
    canonical energy, which is given by
    Eq.~(\ref{eq:canonical_energy_loss}), is identical to the
    corresponding expression in GR if the condition
    (\ref{eq:asymptotic_higgs_condition.vi}) with
    \begin{math}
      \rho_{c} = 0
    \end{math}
    is satisfied. However, when 
    \begin{math}
      \rho_{c} = 0,
    \end{math}
    neither the time average of the emission rate
    for the space-space component of the \lq\lq{}spin'' angular
    momentum
    \begin{math}
      S_{kl}
    \end{math}
    nor that of the orbital angular momentum
    \begin{math}
      L_{[\mu\nu]}
    \end{math}
    are the same as the corresponding expression in GR\@.  The time
    average of the emission rate for the space-space component of the
    sum 
    \begin{math}
      S_{kl} + e^{(0)\mu}_{\hspace{2.8ex}k} 
        e^{(0)\nu}_{\hspace{2.8ex}l} L_{[\mu\nu]}
    \end{math}
    is the same as the corresponding expression in GR\@. However,
    the sum 
    \begin{math}
      S_{kl} + e^{(0)\mu}_{\hspace{2.8ex}k} 
        e^{(0)\nu}_{\hspace{2.8ex}l} L_{[\mu\nu]}
    \end{math}
    is an artificial quantity, because the \lq\lq{}spin'' angular
    momentum $S_{kl}$ and orbital angular momentum $L_{[\mu\nu]}$ are
    associated with transformations that differ with each other.
  \end{enumerate}
 
\item The transformation property of the dynamical energy-momentum
  density of gravity in ENGR is different from that of the canonical
  energy-momentum density of gravity in GR\@. The former behaves as a
  vector density under general coordinate transformations, while the
  latter is not tensorial. Nevertheless, the two densities take the
  same form up to order
  \begin{math}
    O(1 / r^{2}).
  \end{math}

\item In ENGR, the world line of a macroscopic test body is the
  geodesic of the metric $g$, as in the case of GR,\footnote{Note
    that the behavior of the Dirac particle in ENGR differs from that
    in GR\@. (See Ref.~\citen{hayashi.shirafuji:1979}.)}\
  and hence ENGR with the condition (\ref{eq:parameters_conditions})
  accurately describes the variation of the period of motion of the
  binary pulsar PSR1913+16~\cite{hulse.taylor:1975}.
  This is consistent with results in
  Refs.~\citen{schweizer.straumann:1979} and
  \citen{schweizer.straumann.wipf:1980}.
  
\item As far as the gravitational radiation from Newtonian point
  masses treated in the weak-field approximation is concerned, the
  losses of energy-momentum and angular momentum as well as the wave
  form are independent of the parameter $c_{3}$ in the gravitational
  Lagrangian density, and they are identical to the corresponding
  quantities in GR\@. The effects of $c_{3}$ reveal themselves at
  higher orders.
  
\item We have considered the case of a weak field under the condition
  (\ref{eq:parameters_conditions}), but the field equations
  (\ref{eq:weak_symmetric_field_eq}) and
  (\ref{eq:weak_antisymmetric_field_eq}) can be solved also in the
  case with the following~\cite{shirafuji.nashed:1997}:
  \begin{equation}
    \label{eq:parameters_conditions.ii}
    c_{1} + c_{2} \neq 0, \quad
    c_{1} - \frac{4}{9}c_{3} \neq 0.
  \end{equation}
  The dynamical structure of the system with the condition
  (\ref{eq:parameters_conditions.ii}) is significantly different from
  that with the condition (\ref{eq:parameters_conditions}).  Although
  the values of the parameters $c_{1}$ and $c_{2}$ are severely
  restricted as
  \begin{equation}
    c_{1} \simeq -\frac{1}{3\kappa} \simeq - c_{2}
  \end{equation}
  by the results of Solar System
  experiments~\cite{hayashi.shirafuji:1979},
  there still remains the possibility that
  Eq.~(\ref{eq:parameters_conditions}) is not satisfied. The
  gravitational radiation for the case in which the parameters
  $c_{1}$, $c_{2}$ and $c_{3}$ satisfy
  Eq.~(\ref{eq:parameters_conditions.ii}) is also worth examining.
\end{enumerate}

From \ref{discussion:comparing_with_asymptotically_flat_space},
\ref{discussion:choice_of_independent_field_variables} and
\ref{discussion:asymptotic_form_higgs_type_field}, we deduce the
following: The choice 
\begin{math}
  \{\psi^{k}, A^{k}_{\hspace{0.7ex}\mu}, \phi^{A}\}
\end{math}
with Eq.~(\ref{eq:asymptotic_higgs_type_field}) satisfying the
conditions (\ref{eq:asymptotic_higgs_condition.i}) and
(\ref{eq:asymptotic_higgs_condition.ii}) is superior to the other
possible choices, and the generator of \emph{internal}
$\overline{\mbox{Poincar\'{e}}}$ transformations accurately describe
the energy-momentum and angular momentum for a wide class of
gravitating systems, including space-times in which there are
gravitational waves.

In the teleparallel theory of gravity, there have been several 
attempts~\cite{andrade.guillen.pereira:2000,%
maluf.rocha.toribio.branco:2000,maluf.rocha:2001,%
itin:2001,%
blagojevic.vasilic:2000,blagojevic.vasilic:2001} to define
well-behaved energy-momentum and angular momentum densities.  For the
case of the teleparallel equivalent of general relativity, i.e.\
the case with the condition (\ref{eq:teleparallel_equivarence}) in our
notation, this problem is studied in
Refs.~\citen{andrade.guillen.pereira:2000,%
maluf.rocha.toribio.branco:2000,maluf.rocha:2001}.
The gravitational energy-momentum density
\begin{math}
  hj_{a}^{\hspace{0.7ex}\rho}
\end{math}
in Ref.~\citen{andrade.guillen.pereira:2000} is the same as our
\begin{math}
  {}^{G}\bm{T}_{k}^{\hspace{0.7ex}\mu}.
\end{math}
In Refs.~\citen{maluf.rocha.toribio.branco:2000} and
\citen{maluf.rocha:2001}, Hamiltonian formalism is developed, and a
natural definition of the energy-momentum density of the gravitational
field is given. In addition, the angular-momentum density is examined
in Ref.~\citen{maluf.rocha.toribio.branco:2000}. In
Ref.~\citen{itin:2001}, an energy-momentum current that transforms as
a tensor under diffeomorphisms of the space-time manifold and under
global $SO(1,3)$ transformations is proposed in a co-frame field
formulation of the general teleparallel theory of gravity.  In
Refs.~\citen{blagojevic.vasilic:2000} and
\citen{blagojevic.vasilic:2001}, the energy-momentum and angular
momentum densities for general teleparallel theory of an isolated
gravitating system are examined.

The energy-momentum and angular momentum for gravitational waves,
however, are not examined in 
Refs.~\citen{andrade.guillen.pereira:2000,%
maluf.rocha.toribio.branco:2000,maluf.rocha:2001,%
itin:2001,%
blagojevic.vasilic:2000,blagojevic.vasilic:2001}, and in
Refs.~\citen{maluf.rocha.toribio.branco:2000,maluf.rocha:2001,%
itin:2001,%
blagojevic.vasilic:2000,blagojevic.vasilic:2001},
the energy-momentum and angular momentum are not related to the
generators of \emph{internal} Poincar\'{e} transformations, and there
appears no field that corresponds to our $\psi^{k}$.  It is worth
examining the relation between
their~\cite{maluf.rocha.toribio.branco:2000,maluf.rocha:2001,%
itin:2001,%
blagojevic.vasilic:2000,blagojevic.vasilic:2001} energy-momentum and
angular momentum densities and ours.

\section*{Acknowledgements}
The authors would like to express their sincere thanks to Professor
K.~Nakao for valuable discussions.

\appendix
\section{Linearized Einstein Theory}
\label{sec:einstein_theory}

For convenience, we give here a short summary of linearized GR.

The gravitational Lagrangian density is given by
\begin{equation}
  \label{eq:gravitational_lagrangian.GR}
  \bm{L}_{\mathrm{GR}} \stackrel{\mathrm{def}}{=}
  \frac{1}{2\kappa} \sqrt{- g}g^{\mu\nu} \left[%
    {\lambda \brace \mu \hspace{0.7ex} \rho}
    {\rho \brace \nu \hspace{0.7ex} \lambda} -
    {\lambda \brace \mu \hspace{0.7ex} \nu}
    {\rho \brace \lambda \hspace{0.7ex} \rho} \right],
\end{equation}
where we have defined the Christoffel symbols by
\begin{equation}
  \label{eq:chirstoffel_symbol.GR}
  {\lambda \brace \mu \hspace{0.7ex} \nu} \stackrel{\mathrm{def}}{=}
  \frac{1}{2} g^{\lambda\rho} \left(
    \partial_{\mu}g_{\nu\rho} + \partial_{\nu}g_{\rho\mu} -
    \partial_{\rho}g_{\mu\nu} \right).
\end{equation}
The Einstein equation takes the form
\begin{equation}
  \label{eq:einstein_equation}
  R_{\mu\nu}(\{\}) - \frac{1}{2}g_{\mu\nu}R(\{\}) = 
  \kappa T_{\mu\nu} \;,
\end{equation}
where we have defined the Riemann-Christoffel curvature, Ricci tensor
and scalar curvature by
\begin{align}
  \label{riemann_christoffel_curvature}
  R^{\rho}_{\hspace{0.7ex}\sigma\mu\nu}(\{\})
  &\stackrel{\mathrm{def}}{=}
  \partial_{\mu}{%
    \rho \brace \sigma \hspace{0.7ex} \nu} -
  \partial_{\nu}{%
    \rho \brace \sigma \hspace{0.7ex} \mu} +
  {\rho \brace \lambda \hspace{0.7ex} \mu}
  {\lambda \brace \sigma \hspace{0.7ex} \nu} -
  {\rho \brace \lambda \hspace{0.7ex} \nu}
  {\lambda \brace \sigma \hspace{0.7ex} \mu}, \\
  R_{\mu\nu}(\{\}) &\stackrel{\mathrm{def}}{=}
  R^{\rho}_{\hspace{0.7ex}\mu\rho\nu} \;, \\
  R(\{\}) &\stackrel{\mathrm{def}}{=}
  g^{\mu\nu}R_{\mu\nu}(\{\}),
\end{align}
respectively. Also, 
\begin{math}
  T_{\mu\nu}
\end{math}
denotes the energy-momentum density of the gravitational source. The
canonical energy-momentum density is defined by
\begin{equation}
  \label{eq:canonical_emd.GR}
  \tilde{\bm{t}}_{\mu}^{\hspace{0.7ex}\nu} \stackrel{\mathrm{def}}{=}
  \delta_{\mu}^{\hspace{0.7ex}\nu} \bm{L}_{\mathrm{GR}} -
  \frac{\partial \bm{L}_{\mathrm{GR}}}{%
    \partial g_{\rho\sigma,\nu}} g_{\rho\sigma,\mu} \;.
\end{equation}

We consider a metric perturbation
\begin{math}
  h_{\mu\nu}
\end{math}
from Minkowskian space-time, i.e.,
\begin{equation}
  \label{eq:weak_metric.GR}
  g_{\mu\nu} = \eta_{\mu\nu} + h_{\mu\nu} \;, \quad
  |h_{\mu\nu}| \ll 1.
\end{equation}
The linearized field equations are given by the form in
Eq.~(\ref{eq:wave_equation_symmetric}), up to an overall factor of 2,
\begin{equation}
  \label{eq:linearized_field_eq.GR}
  \square \bar{h}_{\mu\nu} = - 2\kappa T_{\mu\nu} \;,
\end{equation}
with the harmonic coordinate condition
\begin{equation}
  \label{eq:harmonic_conditions.GR}
  \partial_{\nu}\bar{h}^{\mu\nu} = 0,
\end{equation}
where we have defined
\begin{align}
    \bar{h}_{\mu\nu} &\stackrel{\mathrm{def}}{=}
    h_{\mu\nu} - \frac{1}{2}\eta_{\mu\nu}h \;, \quad
    h \stackrel{\mathrm{def}}{=} \eta^{\mu\nu}h_{\mu\nu} \;, \\
    \bar{h} &\stackrel{\mathrm{def}}{=}
    \eta^{\mu\nu}\bar{h}_{\mu\nu} \;.
\end{align}
Note that the perturbation
\begin{math}
  \bar{h}_{\mu\nu}
\end{math}
corresponds to
\begin{math}
  2 \bar{f}_{(\mu\nu)}
\end{math}
in \S\ref{sec:weak_field_approximation}. At lowest order,
\begin{math}
  \tilde{\bm{t}}_{\mu}^{\hspace{0.7ex}\nu}
\end{math}
becomes
\begin{align}
  \label{eq:weak_emd_gravity.GR}
  2\kappa \, \tilde{\bm{t}}_{\mu}^{\hspace{0.7ex}\nu} &=
  \delta_{\mu}^{\hspace{0.7ex}\nu} \left(
    \frac{1}{2}\partial^{\rho}\bar{h}^{\lambda\sigma}
    \partial_{\lambda}\bar{h}_{\rho\sigma} -
    \frac{1}{4}\partial^{\lambda}\bar{h}^{\rho\sigma}
    \partial_{\lambda}\bar{h}_{\rho\sigma} +
    \frac{1}{8}\partial^{\sigma}\bar{h}\partial_{\sigma}\bar{h}
  \right) \notag \\
  & \quad +\; \frac{1}{2}\partial_{\mu}\bar{h}_{\rho\sigma}
    \partial^{\nu}\bar{h}^{\rho\sigma} -
    \frac{1}{4}\partial_{\mu}\bar{h}\partial^{\nu}\bar{h} -
    \partial_{\mu}\bar{h}_{\rho\sigma}
    \partial^{\rho}\bar{h}^{\sigma\nu} \;,
\end{align}
which is approximated, up to order
\begin{math}
  O(1 / r^{2}),
\end{math}
as
\begin{equation}
  \label{eq:weak_emd_gravity_leading.GR}
  \tilde{\bm{t}}_{\mu}^{\hspace{0.7ex}\nu} =
  \frac{1}{4\kappa} \left(
  \partial_{\mu}\bar{h}_{\rho\sigma}
    \partial^{\nu}\bar{h}^{\rho\sigma} -
  \frac{1}{2}\partial_{\mu}\bar{h}\partial^{\nu}\bar{h} \right).
\end{equation}

Finally, the energy-momentum density of gravity proposed by Landau and
Lifshitz~\cite{landau.lifshitz} is defined by
\begin{equation}
  \label{eq:symmetric_emd_gravity}
  (- g) t^{\mu\nu}_{\mathrm{LL}} \stackrel{\mathrm{def}}{=}
  \theta^{\mu\nu} - (- g)T^{\mu\nu} \;,
\end{equation}
where 
\begin{math}
  \theta^{\mu\nu}
\end{math}
is the symmetric energy-momentum density defined by
Eq.~(\ref{eq:symmetric_energy_momentum_density.LL}).  By using
Eqs.~(\ref{eq:symmetric_energy_momentum_density.LL}) and
(\ref{eq:einstein_equation}), we obtain from
Eq.~(\ref{eq:symmetric_emd_gravity}) the following:
\begin{align}
  2\kappa \; t^{\mu\nu}_{\mathrm{LL}} &=
  \left(
    2 {\sigma \brace \lambda \hspace{0.7ex} \rho}
      {\tau \brace \sigma \hspace{0.7ex} \tau}
    - {\sigma \brace \lambda \hspace{0.7ex} \tau}
      {\tau \brace \rho \hspace{0.7ex} \sigma}
    - {\sigma \brace \lambda \hspace{0.7ex} \sigma}
      {\tau \brace \rho \hspace{0.7ex} \tau} \right)
  (g^{\mu\lambda}g^{\nu\rho} - g^{\mu\nu}g^{\lambda\rho}) \notag \\
  & \!\! + \, g^{\mu\lambda}g^{\rho\sigma} \left(%
    {\nu \brace \lambda \hspace{0.7ex} \tau}
    {\tau \brace \rho \hspace{0.7ex} \sigma}
    + {\nu \brace \rho \hspace{0.7ex} \sigma}
    {\tau \brace \lambda \hspace{0.7ex} \tau}
    - {\nu \brace \sigma \hspace{0.7ex} \tau}
    {\tau \brace \lambda \hspace{0.7ex} \rho}
    - {\nu \brace \lambda \hspace{0.7ex} \rho}
    {\tau \brace \sigma \hspace{0.7ex} \tau} \right) \notag \\
  & \!\! + \, g^{\nu\lambda}g^{\rho\sigma} \left(%
    {\mu \brace \lambda \hspace{0.7ex} \tau}
    {\tau \brace \rho \hspace{0.7ex} \sigma}
    + {\mu \brace \rho \hspace{0.7ex} \sigma}
    {\tau \brace \lambda \hspace{0.7ex} \tau}
    - {\mu \brace \sigma \hspace{0.7ex} \tau}
    {\tau \brace \lambda \hspace{0.7ex} \rho}
    - {\mu \brace \lambda \hspace{0.7ex} \rho}
    {\tau \brace \sigma \hspace{0.7ex} \tau} \right) \notag \\
  & \!\! + \, g^{\lambda\rho}g^{\sigma\tau} \left(
    {\mu \brace \lambda \hspace{0.7ex} \sigma}
    {\nu \brace \rho \hspace{0.7ex} \tau}
    - {\mu \brace \lambda \hspace{0.7ex} \rho}
    {\nu \brace \sigma \hspace{0.7ex} \tau} \right).
\end{align}
To lowest order, this takes the form
\begin{align}
  2\kappa (- g)t^{\mu\nu}_{\mathrm{LL}} &=
  \eta^{\mu\nu} \left(
    \frac{1}{2} \partial^{\lambda}\bar{h}^{\rho\sigma}
    \partial_{\rho}\bar{h}_{\sigma\lambda} +
    \frac{1}{8} \partial^{\rho}\bar{h}\partial_{\rho}\bar{h} -
    \frac{1}{4} \partial^{\rho}\bar{h}^{\sigma\lambda}
    \partial_{\rho}\bar{h}_{\sigma\lambda} \right) +
    \frac{1}{2}\partial^{\mu}\bar{h}^{\rho\sigma}
    \partial^{\nu}\bar{h}_{\rho\sigma} \notag \\
  & -\,
    \frac{1}{4}\partial^{\mu}\bar{h}\partial^{\nu}\bar{h} -
    \partial^{\mu}\bar{h}_{\rho\sigma}
    \partial^{\rho}\bar{h}^{\nu\sigma} -
    \partial^{\rho}\bar{h}^{\mu\sigma}
    \partial^{\nu}\bar{h}_{\rho\sigma} +
    \partial_{\rho}\bar{h}^{\mu\sigma}
    \partial^{\rho}\bar{h}^{\nu}_{\hspace{0.7ex}\sigma} \;.
\end{align}

\section{Angular Momentum Loss Derived from \\ Energy-Momentum Loss}
\label{sec:problem_three}

Following the argument given in problem 3 in Section 110 of
Ref.~\citen{landau.lifshitz}, we derive the time average of the
orbital angular momentum loss for Newtonian point masses from the time
average of the dynamical energy loss given by
Eq.~(\ref{eq:energy_loss}). The result supports the discussion of the
\lq\lq{}spin'' angular momentum loss given in \S\ref{sec:spin}.

We represent the time average of the energy loss of the system as the
work of the \lq\lq{}frictional forces'' $\bm{\varrho}$ acting on the
Newtonian point masses:
\begin{equation}
  \label{eq:energy_dissipation}
  \left\langle \frac{dE}{dt} \right\rangle =
  \sum_{a = 1}^{N} \left\langle \bm{\varrho}_{a} \cdot
    \dot{\bm{\xi}}_{a} \right\rangle.
\end{equation}
The time average of the loss of angular momentum,
\begin{math}
  \bm{l} \stackrel{\mathrm{def}}{=} 
  \sum_{a = 1}^{N} \bm{\xi}_{a} \times m_{a} \dot{\bm{\xi}}_{a},
\end{math}
is given by
\begin{equation}
  \label{eq:Newtonian_angular_momentum_loss}
  \left\langle \frac{dl_{\alpha}}{dt} \right\rangle =
  \sum_{a = 1}^{N} \left\langle (\bm{\xi}_{a} \times
    \bm{\varrho}_{a})_{\alpha} \right\rangle =
  \sum_{a = 1}^{N} \epsilon_{\alpha\beta\gamma} \left\langle
    \xi_{a}^{\beta}\varrho_{a}^{\gamma} \right\rangle,
\end{equation}
where the symbol 
\begin{math}
  \epsilon_{\alpha\beta\gamma}
\end{math}
denotes the three-dimensional antisymmetric tensor with
\begin{math}
  \epsilon_{123} = \epsilon^{123} = 1.
\end{math}
Note that a Newtonian point mass in ENGR is subject to Newton's
equation of motion. To determine $\bm{\varrho}_{a}$, we write
Eq.~(\ref{eq:energy_loss}) as
\begin{equation}
  \label{eq:modified_energy_loss}
  \left\langle \frac{dE}{dt} \right\rangle =
  - \frac{G}{5c^{5}} \left\langle 
    \dddot{D}_{\alpha\beta}\dddot{D}_{\alpha\beta} -
    \frac{1}{3}\dddot{D}_{\alpha\alpha}\dddot{D}_{\beta\beta}
  \right\rangle =
  - \frac{G}{5c^{5}} \left\langle 
    \dot{D}_{\alpha\beta}{D}^{(\mathrm{v})}_{\alpha\beta} -
    \frac{1}{3}\dot{D}_{\alpha\alpha}{D}^{(\mathrm{v})}_{\beta\beta}
  \right\rangle,
\end{equation}
where we have used the fact that the average values of the total time
derivatives vanish. Here,
\begin{math}
  D^{(\mathrm{v})}_{\alpha\beta}
\end{math}
represents the fifth-order derivative with respect to $t$.
Substituting
\begin{math}
  \dot{D}_{\alpha\beta} = 
  \sum_{a = 1}^{N} m_{a} (\dot{\xi}_{a}^{\alpha}\xi_{a}^{\beta} +
    \xi_{a}^{\alpha}\dot{\xi}_{a}^{\beta})
\end{math}
into Eq.~(\ref{eq:modified_energy_loss}) and comparing with
Eq.~(\ref{eq:energy_dissipation}), we find
\begin{equation}
  \label{eq:explicit_form_frictional_force}
  \varrho_{a}^{\alpha} = - \frac{2G}{5c^{5}} \left(
    D^{(\mathrm{v})}_{\alpha\beta} - 
    \frac{1}{3}\delta^{\alpha\beta}D^{(\mathrm{v})}_{\gamma\gamma}
  \right) m_{a}\xi_{a}^{\beta} \;.
\end{equation}
Substitution of Eq.~(\ref{eq:explicit_form_frictional_force}) into
Eq.~(\ref{eq:Newtonian_angular_momentum_loss}) gives the result
\begin{equation}
  \left\langle \frac{dl_{\alpha}}{dt} \right\rangle = -
  \frac{2G}{5c^{5}} \epsilon_{\alpha\beta\gamma}
  \left\langle \left(
      D^{(\mathrm{v})}_{\gamma\delta} -
      \frac{1}{3}\delta^{\gamma\delta}
        D^{(\mathrm{v})}_{\epsilon\epsilon}\right)
    D_{\beta\delta}
  \right\rangle = -
  \frac{2G}{5c^{5}} \epsilon_{\alpha\beta\gamma}
  \left\langle 
    \ddot{D}_{\beta\delta} \dddot{D}_{\delta\gamma} \right\rangle.
\end{equation}
This is equivalent to Eq.~(\ref{eq:spin_am_space_space_loss}), with
Eq.~(\ref{eq:asymptotic_higgs_type_field}) satisfying the condition
(\ref{eq:asymptotic_higgs_condition.ii}).


\end{document}